\documentclass[footinbib,aps,pra,reprint,twocolumn,superscriptaddress,floatfix]{revtex4-1}
 \usepackage[top=1.4cm,bottom=1.1cm,left=1.90cm,right=1.40cm]{geometry}

\usepackage[utf8]{inputenc}
\usepackage{graphicx}
\usepackage{amsmath}
\usepackage{float}
\usepackage{tabularx}

\begin{document}
	\title{Modeling light shifts in optical lattice clocks}
	\author{Nils Nemitz}
	\affiliation{RIKEN Center for Advanced Photonics, Wako, Saitama 351-0198, Japan}
	\affiliation{Space-Time Standards Laboratory, NICT, Koganei, Tokyo 184-0015, Japan}
	\email[Email: ]{nils.nemitz@nict.go.jp}
	\author{Asbjørn Arvad Jørgensen}
	\altaffiliation[Now ]{Ultra Cold Atoms, The Niels Bohr Institute, University of Copenhagen, Blegdamsvej 17, DK-2100, Denmark}
	\affiliation{Quantum Metrology Laboratory, RIKEN, Wako-shi, Saitama 351-0198, Japan}
	\author{Ryotatsu Yanagimoto}
	\altaffiliation[Now ]{Edward L. Ginzton Laboratory, Stanford University, Stanford, California 94305, USA}
	\affiliation{Quantum Metrology Laboratory, RIKEN, Wako-shi, Saitama 351-0198, Japan}
	\affiliation{Department of Applied Physics, Graduate School of Engineering, The University of Tokyo, Bunkyo-ku, Tokyo 113-8656, Japan}
	\author{Filippo Bregolin}
	\affiliation{Quantum Metrology Laboratory, RIKEN, Wako-shi, Saitama 351-0198, Japan}
	\affiliation{Istituto Nazionale di Ricerca Metrologica (INRIM), Strada delle Cacce 91, 10135 Torino, Italy}
	\affiliation{Dipartimento di Elettronica e Telecomunicazioni, Politecnico di Torino, Corso duca degli Abruzzi 24, 10129 Torino, Italy}
	\author{Hidetoshi Katori}
	\affiliation{RIKEN Center for Advanced Photonics, Wako, Saitama 351-0198, Japan}
	\affiliation{Quantum Metrology Laboratory, RIKEN, Wako-shi, Saitama 351-0198, Japan}
	\affiliation{Department of Applied Physics, Graduate School of Engineering, The University of Tokyo, Bunkyo-ku, Tokyo 113-8656, Japan}
	\date{\today}
\begin{abstract}
We present an extended model for the lattice-induced light shifts of the clock frequency in optical lattice clocks, applicable to a wide range of operating conditions. The model extensions cover radial motional states with sufficient energies to invalidate the harmonic approximation of the confining potential. We re-evaluate lattice-induced light shifts in our Yb optical lattice clock with an uncertainty of $6.1\times10^{-18}$ under typical clock operating conditions.
\end{abstract}

\maketitle
	
\section{Introduction}
Optical frequency standards now reach uncertainties of only a few parts in $10^{18}$~\cite{Ushijima2015, Nicholson2015, Huntemann2016, McGrew2018} by probing narrow transitions of atoms held in strong confinement. For optical lattice clocks, this is achieved by trapping atoms in a large number of lattice sites in the periodic potential of an optical standing wave. The resulting energy shifts of the ground and excited electronic levels are then carefully balanced by tuning the lattice laser to a magic frequency, largely cancelling the resulting shift in the clock transition frequency~\cite{Katori2003}. The degree to which this cancellation can be achieved is limited by frequency shifts that are non-linear in the lattice intensity, associated with electric quadrupole (E2) and magnetic dipole (M1) transitions \cite{Taichenachev2008,Ovsiannikov2013,Ovsiannikov2016} as well as the atomic hyperpolarizability. Achieving a clock uncertainty at the level of $10^{-17}$ or below therefore requires careful evaluation of these effects \cite{Brusch2006, Barber2008, Westergaard2011, Brown2017, Ushijima2018}. 

This evaluation relies on a significant increase in the applied lattice intensity $I$ over what is required for confinement. Besides providing improved leverage, this also separates hyperpolarizability-induced shifts, which scale with $I^2$, from effects that scale as $I$. This results in a design conflict: While a strongly focused trapping beam provides a high available intensity, the tight confinement leads to increased collisional interactions. For the cryogenic optical lattice clocks we have previously reported on \cite{Ushijima2015, Nemitz2016}, the need to create a moving lattice through independent frequency control of the two lattice beams also requires special consideration in the implementation of a resonator-enhanced optical setup, which has elsewhere been successfully implemented \cite{LeTargat2013, Yamanaka2015, Brown2017, Ushijima2018} to alleviate this conflict.

Another concern is evaluation and control of the motional state of the trapped atoms. While clocks operating with ${}^{87}\text{Sr}$ employ effective narrow-line cooling~\cite{Mukaiyama2003} to ensure consistently low thermal energies much smaller than the lattice depth, this is more challenging to achieve for clocks operating with ${}^{171}\text{Yb}$. These observe a degraded cooling efficiency at elevated lattice intensity, which in all likeliness occurs due to significant observed shifts of the $^3P_1$ states by the 759\,nm lattice light.  
Although efficient axial post-cooling is possible by addressing the motional sidebands of the clock transition, controlling radial motion has mostly been realized through rejection of energetic atoms – at the cost of available signal. As a result, Yb clocks typically operate with a large population of atoms with sufficient energy to sample off-axis, peripheral lattice regions where the intensity is significantly reduced. When evaluating the lattice-induced clock frequency shifts, it is essential to separate these experiment-specific properties from the underlying physical quantities, if the results are to be tested by other clocks \cite{Zhang2015, Pizzocaro2017, Kim2017}, or applied to different optical configurations.
   
Here we present an amended light shift model based on a description of trapping conditions through parameters available from spectroscopic data. This allows measuring the relevant coefficients using a configuration modified for increased intensity, and applying the results directly to the nominal clock configuration. For typical operating conditions of our cryogenic Yb optical lattice clock, the improved light shift evaluation yields an uncertainty of $6.1\times10^{-18}$, a five-fold reduction from the previously published value \cite{Nemitz2016}.

\section{Measurements}
\begin{figure}[!tp]
	\includegraphics[width=8.35 cm]{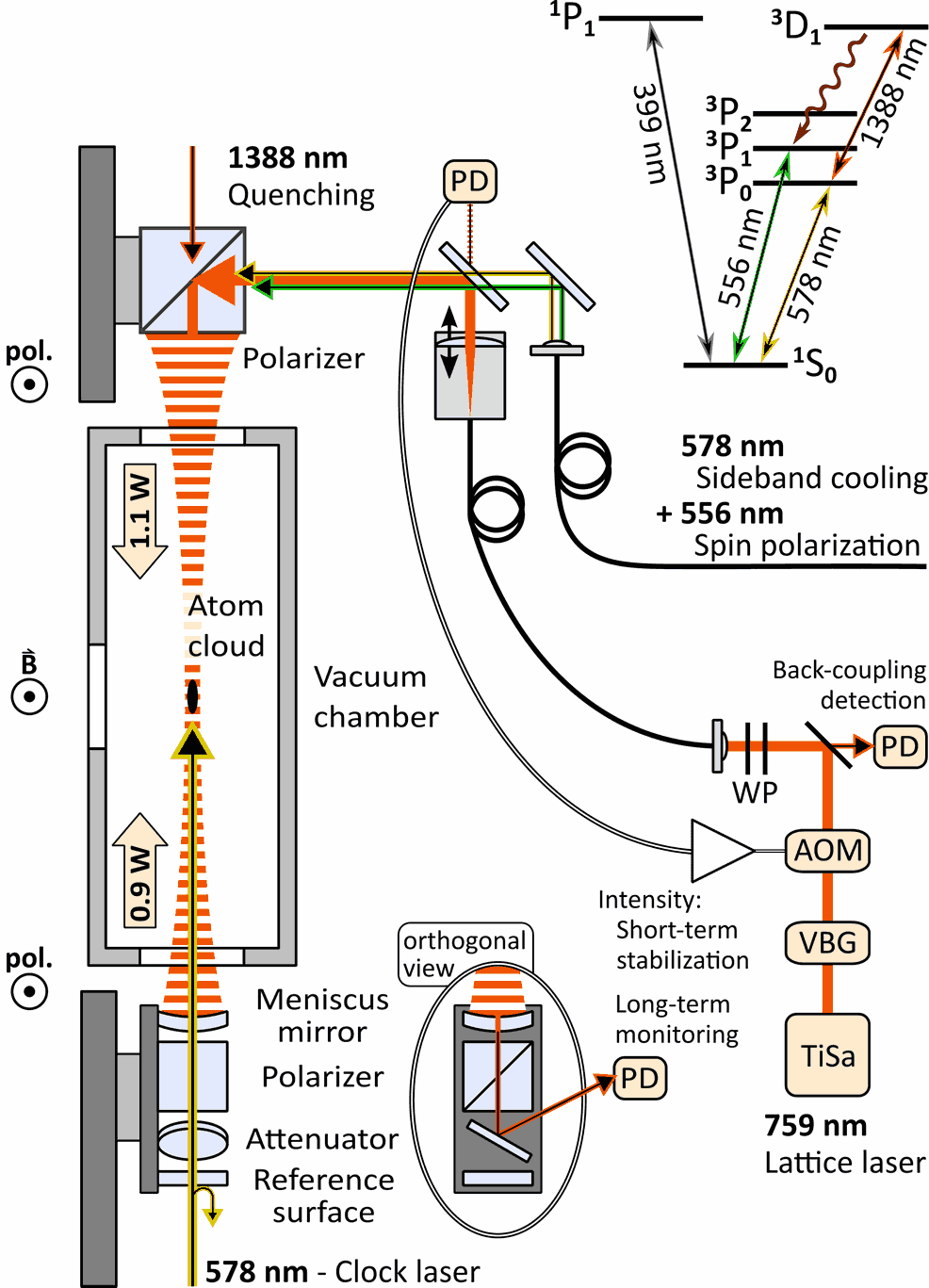}
	\caption{Optical  setup for light shift measurements. The lattice is formed by a retro-reflected focused beam at 759\,nm. The meniscus shape of the retro-reflecting end mirror allows the clock laser at 578\,nm to pass nearly undistorted, and an additional flat surface serves as reference for Doppler-noise cancellation. Lasers at 556\,nm, 578\,nm and 1388\,nm are superimposed on the lattice beam. Photodiodes (PD) monitor the lattice beam intensity before and after the chamber, for fast stabilization through feedback to an acousto-optic modulator (AOM) and long-term monitoring. The output of the Ti:sapphire laser source (TiSa) is spectrally filtered by a volume Bragg grating (VBG) with a FWHM of 40\,GHz, while waveplates (WP) before the fiber coupler allow correction for birefringent effects. The retro-reflected light recoupled into the optical fiber is detected to optimize beam overlap through adjustments to retro-mirror and focal position of the original beam. Inset shows relevant transitions in Yb.}
	\label{schematic}
\end{figure}

For the current experiments, the experimental setup has been equipped with a retro-reflected lattice with reduced beam radius (see Fig.~\ref{schematic}). The lattice light is transported to the chamber by a 1\,m long end-capped polarization-maintaining optical fiber that supplies a beam with maximum power of 1.1\,W to the atoms. In the following, we will use the depth $V_0$ of the sinusoidal on-axis trapping potential to indicate the lattice standing-wave intensity, since this is directly accessible to spectroscopic measurements through the axial trap frequency $f_z = 2 \sqrt{V_0  E_r} / h$. The lattice photon recoil energy $E_r=(h \, \nu_\text{E1} )^2 / (2 \, m_\mathrm{Yb} \, c^2) = h\times 2.02\mathrm{\,kHz}$ is also used as a convenient unit throughout the paper. Using a theoretical value of $\alpha_\mathrm{E1}\approx h \times 8.7\mathrm{\,kHz/(kW/cm^2)}$ \cite{Dzuba2010} for the E1 polarizability at the magic frequency $\nu_\mathrm{E1}\approx394.8\mathrm{\,THz}$, the observed depth $V_0=650 \,E_r\approx k_\mathrm{B}\times 63 \mathrm{\,\mu K}$ corresponds to a maximum intensity of $I_0\sim 150 \mathrm{\,kW/cm^2}$ at the lattice anti-nodes, consistent with a beam radius of $w\sim 43 \mathrm{\,\mu m}$ at the trap position.

The lattice light is reflected back on itself by a meniscus-shaped mirror with a concave radius of curvature of 93\,mm, coated for high reflection at 759\,nm and high transmission at 556--578\,nm and 1388\,nm. To avoid Doppler shifts, the 578\,nm clock laser ($\nu_\mathrm{Yb}= 518.295\,837\,\mathrm{THz}$) used for interrogation is phase-stabilized to a flat reference surface ($R=0.9$) mounted to the same structure as the retro-mirror that determines the location of the lattice anti-nodes and thereby the trapped atoms. An additional partially transmitting mirror ($T=0.01$) further attenuates the beam to allow for $\pi$-pulses of 60 to 300\,ms length. The shape of the retro-mirror retains the collimation of the clock laser beam. A dichroic mirror in the top path of the lattice laser admits unattenuated beams at 556\,nm and 578\,nm for state preparation and characterization. Polarizing beam splitters (PBS) ensure a common polarization axis for all beams. The magnetic field is aligned to the same axis during preparation and interrogation of the atomic sample. A beam at 1388\,nm is superimposed on the lattice using the out-of-band transmission of the lattice PBS and retains both parallel and orthogonal polarization components. This allows for frequency-selective excitation of a specific Zeeman component of the ${}^3D_1 (F=\tfrac{1}{2})$ state used for quenching the ${}^3P_0$ state during sideband cooling~\cite{Nemitz2016}, and is used to assist spin-polarization: By populating a component that decays to the ${}^1S_0$ ground state with branching ratios favoring the desired Zeeman component, the pumping cycles on the ${}^1S_0 \to {}^3P_1 (F=\tfrac{3}{2})$  transition required to create a spin-polarized sample are minimized. With sideband cooling and spin pumping applied either simultaneously or as a sequence of alternating pulses, we typically achieve 98\,\% spin-polarization at an average vibrational quantum state of $\bar{n}<0.1$. 

\begin{figure}[!tbp]
	\includegraphics[width=8.7 cm]{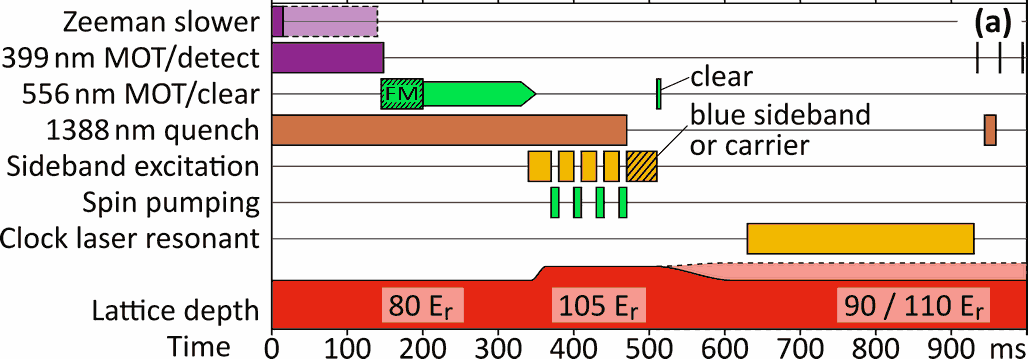}\\ 
	\vskip 2mm 
	\includegraphics[width=8.9 cm]{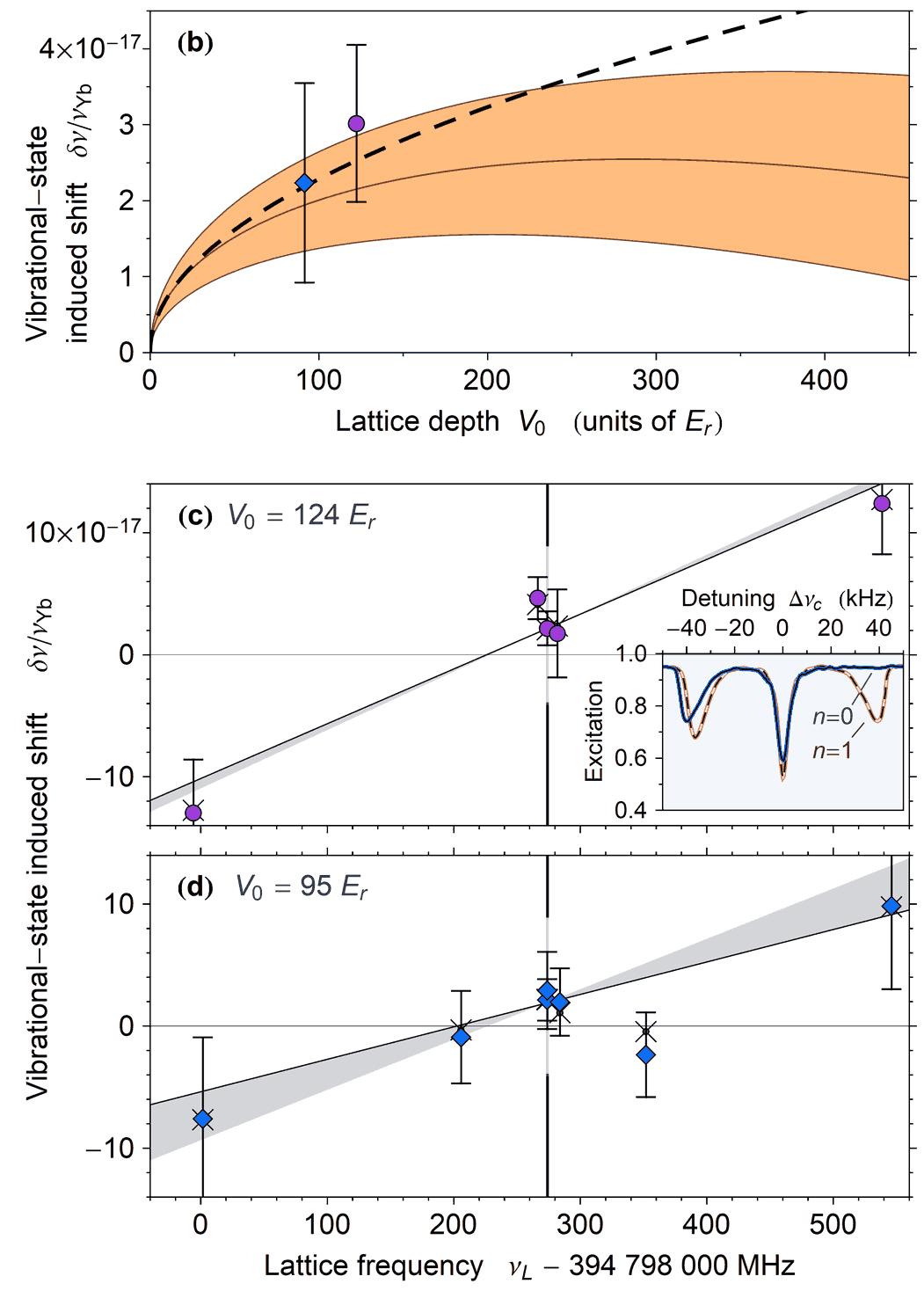}
	\caption{Frequency shifts with axial vibrational state. (a) Experimental sequence. A clock transition pulse (hatched) of $\gg \pi$ area, resonant with either carrier or blue sideband, is applied to a spin-polarized ensemble with $\bar{n}\approx 0$. After a ${}^1S_0 \leftrightarrow {}^3P_1$ clearing pulse only atoms in ${}^3P_0$ remain, occupying either $n=0$ or $n=1$. Clock operation evaluates atoms returned to ${}^1S_0$. (b) Differential light shift between $[n=1]$ and $[n=0]$ as a function of lattice depth. Dashed line represents $\left[ \tilde{\alpha}' \left( \nu_\mathrm{L} - \nu_\mathrm{E1} \right) - \tilde{\alpha}^{qm} \right] \left(\bar{n}+\tfrac{1}{2} \right)\sqrt{\left( \zeta-\tfrac{1}{2}\delta_2 \right) \tfrac{V_0}{E_r}}$, the dominant contribution for small $V_0$. For larger depths, interaction terms incorporating $\bar{n}$ and $\tilde{\beta}$ become relevant. Solid line includes all model terms, with shaded region indicating $1\sigma$ confidence band for $\tilde{\alpha}^{qm}/h=-1027(378)\,\mathrm{\mu Hz}$.
Data points show weighted means of experiments taken near $\nu_\mathrm{E1}$. (c) Full data for differential light shift between $[n=1]$ and $[n=0]$ at $V_0 = 124\,E_r$. Solid line represents model predictions, gray-shaded area indicates correction for $\zeta_{[n=1]}= 0.774 > \zeta_{[n=0]}= 0.768$.
Results before collisional corrections are marked $\times$, interrupted vertical line shows $\nu_\mathrm{L}$ for model of (b). Inset shows sideband spectra for $[n=1]$ and $[n=0]$. Even if $n=1$ for all atoms, the resulting spectrum retains an asymmetry due to different Rabi frequencies for red and blue sidebands (Appendix~\ref{sec_sideband}).
(d) Differential light shift at $V_0 = 95\,E_r$. Solid line represents model predictions, gray-shaded area indicating correction for $\zeta_{[n=1]}= 0.789 > \zeta_{[n=0]}= 0.754$.
}
	\label{multipolar}
\end{figure}

While the state-preparation sequence varies between different series of measurements, as shown in later Figures \ref{hyperpolar1}(a), \ref{hyperpolar2}(a) and \ref{multipolar}(a), it is maintained for all experiments in a series. To explore lattice-intensity induced frequency shifts, the lattice depth is adiabatically ramped to the desired value only after state preparation has been completed.

All measurements described here are performed in interleaved operation, with the clock alternating between three or four distinct measurement conditions typically varying in lattice depth and atom number. The clock transition frequency relative to the frequency of the cavity-stabilized laser is tracked for each condition. The frequency differences between these independent trackers correspond to the systematic shifts due to the change in operating conditions and are insensitive to common effects such as AC/DC Stark shifts from blackbody radiation and parasitic charges.

\begin{figure}[!tbp]
	\includegraphics[width=8.9 cm]{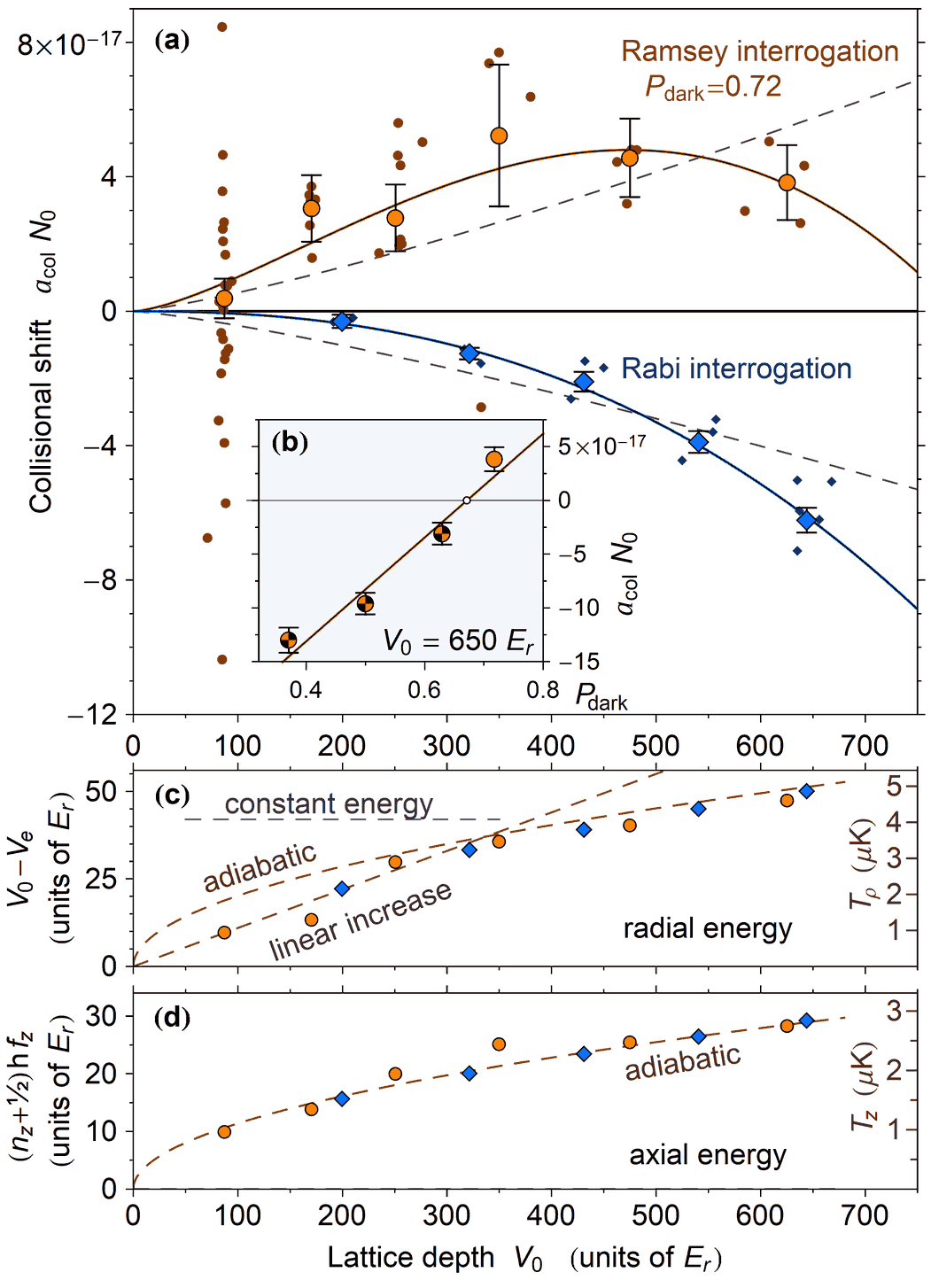} 
	\caption{(a) Collisional shifts at nominal atom number $N_0$ as function of lattice depth. Measurements alternating high and low atom numbers ($N_h$ and $N_l$) yield a normalized collisional shift coefficient $a_\mathrm{col}=(\delta\nu_\mathrm{col}/\nu_\mathrm{Yb})/(N_h - N_l)$ with strong dependence on $V_0$. Markers with error bars show binned results, based on individual data points (without error bars for clarity) obtained during the light shift measurements, and using either Rabi or Ramsey interrogation (as labeled). The apparent scatter at the light shift reference condition $V_0\approx90\,E_r$ results from the large number of measurements. Ramsey interrogation with initial excitation to $P_\mathrm{dark}=0.72$ results in a partial cancellation of collisional shifts for large $V_0$, increased uncertainties are due to operation at reduced atom number.
Dashed lines show a fit proportional to
$V_0^{5/4}$, expected for the scaling of $p$-wave interactions with the observed adiabatic change of energy with confinement. A model scaling with $V_0^{3/2}$ (representing constant thermal energy) also fails to match the data. Solid lines serve to guide the eye and represent a minimal empirical model $c_1 V_0^{5/2}+c_2 V_0^{3/2}$ that adds a higher order term.
(b) Collisional shift cancellation. Measurements of $a_\mathrm{col}$ show a linear dependence on $P_\mathrm{dark}$, consistent with a significant $p$-wave contribution. Plain, orange marker shows data taken at operating point $P_\mathrm{dark}=0.72$ over the course of the light shift experiments. Full data set finds cancellation of collisional shifts at $P_c=0.67$ for $V_0=650\,E_r$. 
(c) After an initial linear increase, relative radial potential energy $(V_0-V_e)$ (see Appendix \ref{sec_sideband}) remains proportional to $\sqrt{V_0}$ over a large range of lattice depth. Right hand axis indicates radial temperature $T_\rho = (V_0-V_e) / k_B$. Dashed lines illustrate scaling for different thermal models.
(d) Axial vibrational state remains $\bar{n}\approx 0.1$ after initial sideband cooling, to give axial motional energy $W_z = (\bar{n}+\tfrac{1}{2}) h f_z$. Right hand axis shows axial temperature $T_z = W_z/ k_B$.}
	\label{collisions}
\end{figure}

The same is not true for atomic interactions, which result in collisional frequency shifts that vary with confinement and atomic temperature. For the initial measurement series using Rabi interrogation, these exceed $-5\times 10^{-17}$ at the largest investigated lattice depths, despite limiting the number of trapped atoms to approximately $N_0=250$, distributed among a similar number of lattice sites. To separate lattice light shifts from collisional shifts, we extrapolate all results to zero density. For the majority of measurements (including all measurements performed at increased lattice intensity), this extrapolation is based on additional interleaved measurements performed at an atom number elevated to three times the normal value. The simultaneous measurement of the collisional shift avoids assumptions on the long term stability of trapping conditions, and allows treating the resulting uncertainty as statistical in nature. Figure~\ref{collisions}(a) and (b) summarize the observed collisional shifts. We rely on an interpolation model only at low lattice depth, where the typical magnitude of the collisional shifts is $1.5\times 10^{-18}$.

To confirm the extrapolation to zero collisional shift, we change the experimental conditions during a second measurement series. Here we use Ramsey excitation with an excitation probability $P_\mathrm{dark}=0.72$ during the dark time to reduce and reverse the frequency shifts resulting from atomic interactions \cite{Ludlow2011, Yanagimoto2018}. This is realized as a sequence consisting of two clock laser pulses with lengths $\tau_1=38.6\mathrm{\,ms}$ and $\tau_2=21.4\mathrm{\,ms}$, for a combined pulse area $2 \pi\times f_c (\tau_1+\tau_2)=\pi$ at a clock transition Rabi frequency $f_c$. These are separated by a 150\,ms dark period during which the clock laser is detuned by 200\,kHz and attenuated to 10\,\% intensity to minimize interaction with the atoms while maintaining phase stabilization to the reference surface. This sequence results in a reversal of the collisional shifts, with a maximum observed shift of $+1.3\times 10^{-17}$ with close to 75 atoms, as typically used throughout this series. Under such conditions approximately 26\,\% of the atoms are expected to reside in at least doubly populated sites that allow interactions. In a series of experiments reported elsewhere~\cite{Yanagimoto2018}, we find no evidence for a non-linearity of the collisional shifts in terms of detected atom number that might compromise the extrapolation to zero density. 

A theoretical model based on \cite{Kanjilal2004} and \cite{Lemke2011} predicts a scaling of the dominant $p$-wave contribution $\propto V_0^{5/4}$ if the distribution across vibrational levels remains constant during changes in confinement, corresponding to an adiabatic change in temperature. If instead the ensemble motional energy 
is maintained despite changes in confinement, the atomic interactions scale $\propto V_0^{3/2}$ \cite{Nicholson2015}, reflecting available volume. 

We control the axial vibrational state through sideband cooling and confirm $\bar{n} \approx 0.1$ after the ramp to final trapping conditions. As elaborated in Appendix~\ref{sec_sideband}, the effective lattice depth $V_e$, discussed in the following section, provides information on the radial potential energy. Except at the lowest lattice depths, we find $V_0-V_e$ to scale with $\sqrt{V_0}$, as expected for unchanged vibrational quantum numbers in a potential with radial trap frequency $f_\rho = \sqrt{V_0 / m_\mathrm{Yb}} / (\pi w)$. Figures~\ref{collisions} (c) and (d) illustrate the radial and axial energies with lattice depth.  

While this supports collisional effects scaling with $V_0^{5/4}$, our experiments indicate the presence of a higher order term with a negative sign for both Rabi and Ramsey measurements. We attribute this to interactions with atoms returned to the ground state by off-resonant scattering of lattice photons \cite{Doerscher2018}. As such scattering predominantly occurs from the excited ${}^3P_0$ state, the additional collisional contribution increases not only with lattice intensity, but also with initial excitation $P_\mathrm{dark}$, consistent with the observed shift of the collisional cancellation point away from the expectation of $P_c\approx 0.5$~\cite{Ludlow2011}. For high lattice intensities, the population of non-coherent ground state atoms makes up more than one percent of the total atom number. Further investigation will be needed to develop a complete model.

\subsection{Light shift model}

After accounting for collisional interactions, we evaluate the clock frequency shifts with varying intensity at different lattice frequencies $\nu_\mathrm{L}$ within the model framework already used in \cite{Nemitz2016} and \cite{Ushijima2018}. For two counterpropagating plane waves of equal intensity, this describes the lattice-light induced shift $\Delta\nu_i$ for a trapped atom in vibrational state $n_i$ as
\begin{equation}\label{eq_plane_wave}
\begin{split}
h \, \Delta\nu_i &= \left[ \tilde{\alpha}' \left( \nu_\mathrm{L} - \nu_\mathrm{E1} \right) - \tilde{\alpha}^{qm} \right] \left(n_i+\tfrac{1}{2} \right) \sqrt{\tfrac{V}{E_r}}\\
                 &-\left[ \tilde{\alpha}' \left( \nu_\mathrm{L} - \nu_\mathrm{E1} \right) + \tfrac{3}{4}\tilde{\beta} \left( 2n_i^2+2n_i+1 \right) \right] \tfrac{V}{E_r}\\
                 &+\tilde{\beta} \left( 2n_i+1 \right) \left( \tfrac{V}{E_r} \right)^{3/2} 
                    -\tilde{\beta} \left( \tfrac{V}{E_r} \right)^2 \quad,
\end{split}
\end{equation}
where $V$ is the depth of the resulting sinusoidal lattice potential~\cite{Katori_Ovsiannikov2015}. The coefficients 
\begin{equation}
\begin{split}
    \tilde{\alpha}' &\equiv \tfrac{\partial}{\partial \nu_\mathrm{L}} \Delta{\alpha}_\mathrm{E1} \left( E_r / \alpha_\mathrm{E1} \right) \\
    \tilde{\alpha}^{qm}&\equiv \Delta{\alpha}^{qm} \left( E_r / \alpha_\mathrm{E1} \right) \\
    &\equiv \left( \Delta{\alpha}_\mathrm{M1} + \Delta{\alpha}_\mathrm{E2} \right) \left( E_r/ \alpha_\mathrm{E1} \right) \\
\text{and}\quad \tilde{\beta} &\equiv \Delta{\beta} \left( E_r / \alpha_\mathrm{E1} \right)^2 
\end{split}
\end{equation}
respectively describe the shifts resulting from the slope of the differential E1 polarizability $\Delta{\alpha}_\mathrm{E1}$ around the E1 magic frequency $\nu_\mathrm{E1}$, the combined differential M1 and E2 polarizabilities $\Delta{\alpha}_\mathrm{M1}$ and $\Delta{\alpha}_\mathrm{E2}$, as well as the differential hyperpolarizability $\Delta{\beta}$. 

To apply this equation to a lattice with a finite beam radius and populated by multiple atoms, we include radial atomic motion by considering the different powers $V^m$ (where $m$ is one of the exponents $\tfrac{1}{2}$, $1$, $\tfrac{3}{2}$ or $2$) as averages over atomic trajectories for the entire ensemble. For a suitably large number of atoms or experimental repetitions, this can be expressed as an effective value
\begin{equation}\label{eq_distribution}
V_e^m\approx \int{ V^m \sigma \left( V \right)\, dV } \quad .
\end{equation}
The distribution $\sigma (V)$ expresses the probability that at a given instant, a randomly chosen atom occupies a position with an axial, sinusoidal depth $V$ (e.g. an off-axis location with reduced intensity). 
Conveniently, $\sigma (V)$ is experimentally accessible through sideband spectroscopy. Including a quartic correction for the axial anharmonicity of the potential, the blue-sideband transition $n_i \to n_i+1$ occurs at a detuning of 
\begin{equation}\label{eq_delta_BSB}
\delta^B(V) = f_z - \frac{(n_i+1) E_r}{h} = \frac{ 2 \sqrt{V E_r} - (n_i+1) E_r }{ h }
\end{equation}
relative to the $n_i \to n_i$ carrier transition \cite{Blatt2009}. Sideband spectra are acquired by applying high intensity clock laser pulses of $1\,\mathrm{ms}$ duration through the same unattenuated path used for sideband cooling (Fig.~\ref{schematic}). By numerical optimization, we find a set of $V_i$ that provides a discretized approximation $\sigma_d(V)$ and reproduces the shape of the blue sideband, as discussed in Appendix \ref{sec_sideband}. We take the largest $V_i$ within the set to represent $V_0$.

This approach yields $\sigma(V)$ without relying on approximations for the shape of the radial potential and the atomic energy distribution, which is of particular importance when it is not possible to cool atoms to radial motional energies $W_\rho \ll V_0$. In the presence of a population of barely trapped atoms with energies approaching $V_0$, the model of \cite{Blatt2009}, which assumes a thermal energy distribution in a harmonic radial potential, fails to reproduce the features of the sideband spectra, as seen in Fig~\ref{radial}.

A limitation to the direct extraction of $\sigma(V)$ is that it requires effective axial sideband cooling to $n_i=0$ to resolve ambiguities of Eq.~\ref{eq_delta_BSB}. We typically observe $>90\,\%$ population of $n=0$.
The residual population of excited state atoms is included in the calculated spectrum as an $n=1$ population that experiences the same distribution $\sigma(V)$.

To describe the trapping conditions with a minimal set of parameters, we define a fractional depth $\zeta$ based on Eq.~\ref{eq_distribution} that relates the effective depth $V_e$, averaged across the atomic ensemble, to the maximal, on-axis lattice potential depth $V_0$ as
\begin{equation}
\label{eq_effective_depth}
V_e \approx \zeta V_0 = \int{ V \sigma \left( V \right)\, dV } \quad .
\end{equation}
Small values of $\zeta$ represent energetic ensembles, where atoms deviate further from the lattice axis.  
A set of small corrections $\delta_{1/2}$, $\delta_{3/2}$ and $\delta_2$ accommodate averaging over the respective powers $m \neq 1$ of the lattice depth in Eq.~\ref{eq_plane_wave}:
\begin{equation}
V_e^m \approx \left[ \left( \zeta+\delta_m \right) V_0 \right]^m = \int{ V^m \sigma \left( V \right)\, dV }
      ,\ m = \tfrac{1}{2},\tfrac{3}{2},2
\end{equation}
These corrections gain significance when $\sigma(V)$ is non-zero over a large range of $V$. Although all $\delta_m$ are directly available from $\sigma(V)$, we find it convenient to eliminate $\delta_{1/2}$ and $\delta_{3/2}$ as independent parameters by expressing them as $\delta_{1/2} \approx -\tfrac{1}{2} \delta_2$ and $\delta_{3/2} \approx \tfrac{1}{2} \delta_2$. These relations match the numerical results and agree with analytical calculations for polynomial potentials up to fourth order in radial position $\rho$.

For the frequency shift $\Delta\nu_\mathrm{en}$ observed over the entire ensemble of atoms in varying motional states, Eq.~\ref{eq_plane_wave} then takes on the form
\begin{equation}
\begin{split}\label{eq_ensemble}
h \, \Delta\nu_\mathrm{en} &= \left[ \tilde{\alpha}' \left( \nu_\mathrm{L} - \nu_\mathrm{E1} \right) - \tilde{\alpha}^{qm} \right] \left(\bar{n}+\tfrac{1}{2} \right) 
                     \sqrt{\left( \zeta-\tfrac{1}{2}\delta_2 \right) \tfrac{V_0}{E_r}}\\
       &-\left[ \tilde{\alpha}' \left( \nu_\mathrm{L} - \nu_\mathrm{E1} \right) + \tfrac{3}{4}\tilde{\beta} \left( 2\bar{n}^2+2\bar{n}+1 \right) \right] \zeta \tfrac{V_0}{E_r}\\
       &+\tilde{\beta} \left( 2\bar{n}+1 \right) \left[ \left( \zeta+\tfrac{1}{2}\delta_2 \right)\tfrac{V_0}{E_r} \right]^{3/2} \\
       &-\tilde{\beta} \left[ \left( \zeta+\delta_2 \right) \tfrac{V_0}{E_r} \right]^2 \quad,
\end{split}
\end{equation}
where trapping conditions for any ensemble of trapped atoms are described by the parameters $V_0$, $\zeta$ and $\delta_2$, in addition to the lattice frequency $\nu_\mathrm{L}$ and the mean axial vibrational state $\bar{n}$. The single $\bar{n}^2$ term is sufficiently small not to require a correction as long as the variation of $n_i$ across the ensemble is controlled by sideband cooling. Parameter values for a range of conditions are given in Fig.~\ref{hyperpolar2}. A similar analysis for Sr \footnote{Note that we use a slightly different notation here, such that $(\zeta + \delta_m)^m = \zeta_m^\mathrm{Sr}$} finds $\zeta_\mathrm{Sr} \approx 0.9$, and insignificantly small corrections $\delta_m$ \cite{Ushijima2018}. 

\begin{figure}[!bph]
	\includegraphics[width= 8.3 cm]{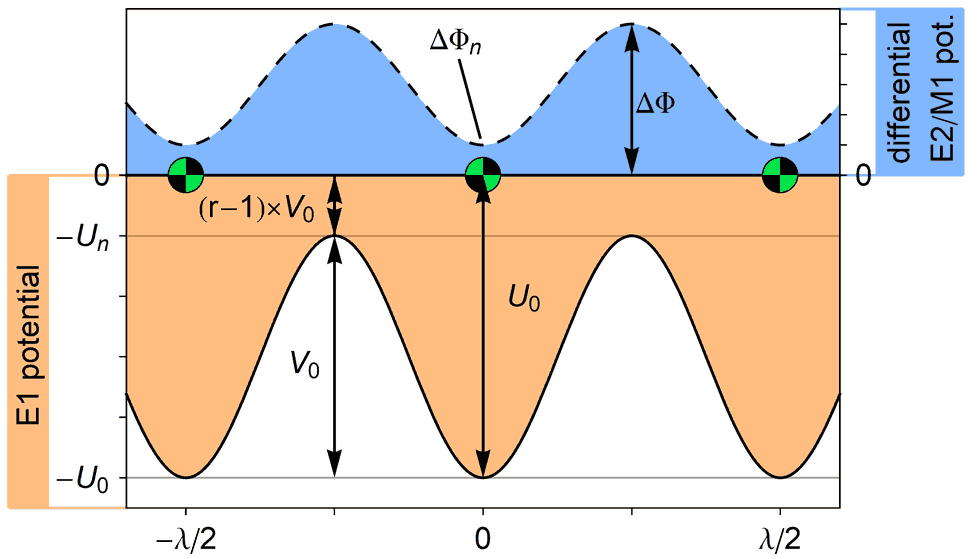}
	\caption{Potential depth. Solid, orange-shaded curve shows lattice E1 potential in the presence of an intensity imbalance between lattice beams. Recovering the total potential depth $U_0$  then requires a correction $r = U_0 / V_0$ to the smaller depth of modulation $V_0$ extracted from sideband spectra. The same reduction in modulation affects the differential E2/M1 potential, which appears phase-shifted by $\lambda/4$ (dashed, blue-shaded curve) and for intensity-balanced beams has a zero-valued node at the trapping positions (circular markers). In the presence of an intensity imbalance, the residual $\Delta{\Phi}_n$ at the trapping positions gives rise to additional clock frequency shifts. Note that the upper vertical axis has been greatly expanded to show $\Delta{\Phi}\ll U_0$.}
	\label{potentials}
\end{figure}

A concern for the precise determination of the light shift coefficients is that an imbalance in the lattice beam intensities may introduce a running-wave contribution. As illustrated in Fig.~\ref{potentials}, it is then necessary to distinguish between the sinusoidal depth of modulation $V_0$, as probed by sideband spectroscopy, and the total potential depth $U_0 = \alpha_\textrm{E1} I_0$ that directly corresponds to intensity. We define a factor $r = U_0 / V_0 \ge 1$ to incorporate this distinction in the light shift model. A secondary aspect of a running wave contribution is the residual potential $U_n = (r-1) V_0$ at the former lattice nodes. The potential resulting from the E2 and M1 polarizabilities is affected in the same way: Instead of a zero-valued node at the trap position, an intensity imbalance yields a differential nodal potential $\Delta{\Phi}_n = -(r-1)  V_0 \, \Delta{\alpha}^{qm} / \alpha_\mathrm{E1}$. This gives rise to an additional term of $-(r-1)  \tilde{\alpha}^{qm} \, V_0 / E_r$, such that the overall light shift equation takes on the form 
\begin{widetext}
\begin{equation}
\begin{split}\label{eq_light_shift}
h \, \Delta\nu_\mathrm{ls} &= \left[ \tilde{\alpha}' \left( \nu_\mathrm{L} - \nu_\mathrm{E1} \right) - \tilde{\alpha}^{qm} \right] \left(\bar{n}+\tfrac{1}{2} \right) 
                     \sqrt{\left( \zeta-\tfrac{1}{2}\delta_2 \right) \tfrac{V_0}{E_r}}\\
       &-\left[ \tilde{\alpha}' \left( \nu_\mathrm{L} - \nu_\mathrm{E1} \right) r 
             + \tilde{\alpha}^{qm} (r-1)
             + \tfrac{3}{4}\tilde{\beta} \left( 2\bar{n}^2+2\bar{n}+1 \right) \right] \zeta \tfrac{V_0}{E_r}\\
       &+\tilde{\beta} \left( 2\bar{n}+1 \right) r \left[ \left( \zeta+\tfrac{1}{2}\delta_2 \right)\tfrac{V_0}{E_r} \right]^{3/2}
       -\tilde{\beta} \left[ r \left( \zeta+\delta_2 \right) \tfrac{V_0}{E_r} \right]^2 \quad.
\end{split}
\end{equation}
\end{widetext}
Terms incorporating the vibrational state $\bar{n}$ represent the finite extent of the atomic waveforms sampling the curvature of the potential, which corresponds directly to $V_0$. Therefore, no factor $r$ appears here, with the exception of the $(V_0 / E_r)^{3/2}$ term, which includes $r$ due to the quadratic intensity dependence of the hyperpolarizability. The added E2/M1 term manifests as a (typically negligible) offset to the extracted E1 magic frequency $\nu_\mathrm{E1}$, as can be seen by combining the terms linear in $V_0$ to find a new apparent value 
$\widetilde{\nu} = \nu_\mathrm{E1} - \left( \tilde{\alpha}^{qm} / \tilde{\alpha}' \right) \left( 1-\tfrac{1}{r} \right)$.

\subsection{Determination of $\tilde{\alpha}'$, $\tilde{\beta}$ and $\nu_\textrm{E1}$}
 
\begin{figure}[!tbp]
	\includegraphics[width=8.7 cm]{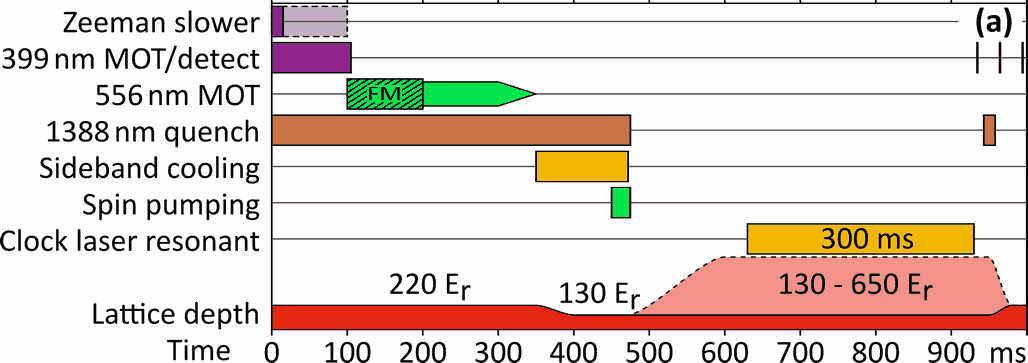}\\
	\vskip 2mm 
	\includegraphics[width=8.9 cm]{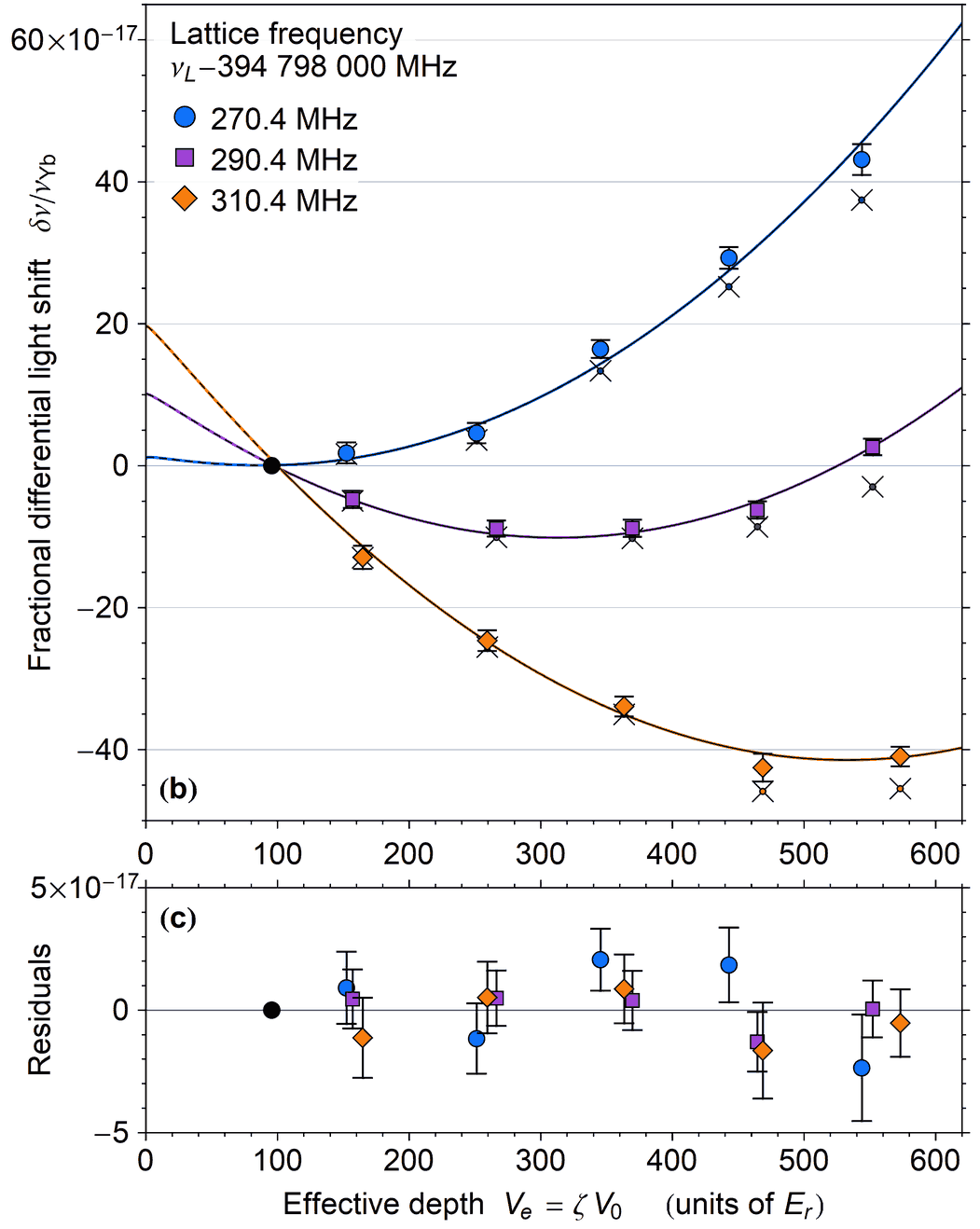}
	\caption{Measurement of intensity-dependent light shifts using Rabi interrogation. (a) Experimental sequence. Lattice depth shown as $V_0$, with lightly shaded region indicating explored range of trapping conditions. During the period marked ‘FM’, the 556\,nm trapping laser is broadened by frequency modulation. Initial operation of the Zeeman slower beam is varied to control the trapped atom number. (b) Differential light shift between a high intensity test condition at an effective lattice depth (see Eq.~\ref{eq_effective_depth}) of $V_e = 150 \textrm{ to } 570\,E_r$  and a reference condition at $V_e=95\,E_r$ (indicated by black anchor point), clearly resolving the hyperpolarizability. Error bars indicate $1\sigma$ uncertainty after accounting for collisional shifts and variation of trapping parameters. Markers and colors indicate lattice frequencies. Additional markers $\times$ show results before correction for collisional effects. Lines show the predictions of the light shift model (Eq.~\ref{eq_light_shift}) for interpolated trapping parameters, intersecting under reference conditions. (c) Deviation of the measured results from the model for the specific trapping parameters of each measurement, with no interpolation of trapping parameters.}
	\label{hyperpolar1}
\end{figure}

\begin{figure}[!tbp] 
	\includegraphics[width=8.7 cm]{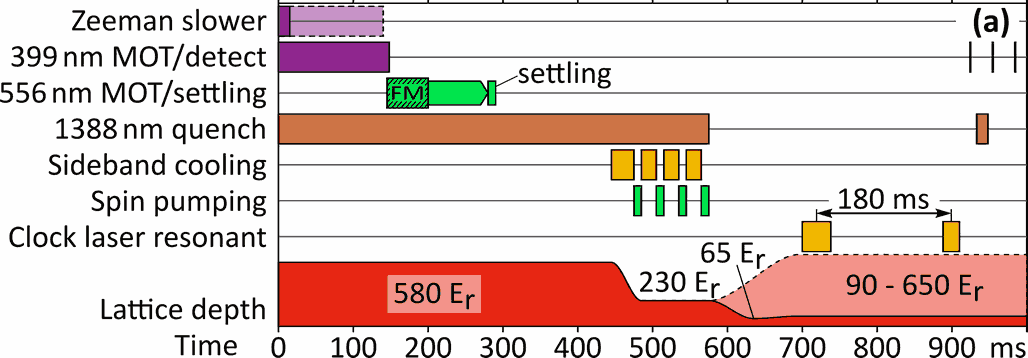}\\
	\vskip 2mm 
	\includegraphics[width=8.9 cm]{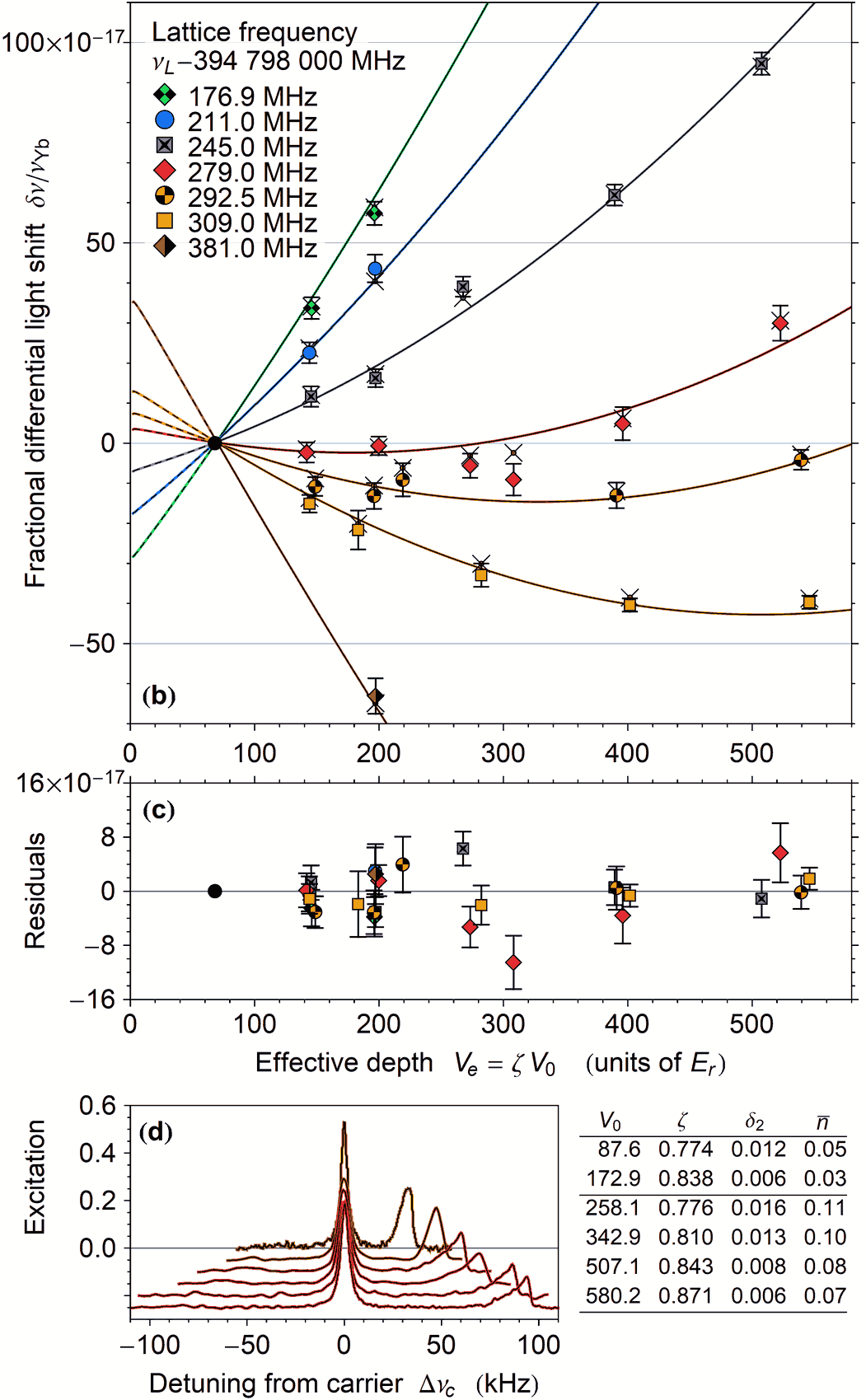}
	\caption{Measurement of intensity-dependent light shifts using Ramsey interrogation. (a) Experimental sequence. Lattice depth shown as $V_0$, lightly shaded region indicates explored range of trapping conditions. During ‘settling’, a pair of independently-controlled radial beams at 556\,nm provides additional Doppler cooling. For final $V_0<200\,E_r$, a temporary reduction to $65\,E_r$ lattice depth expels weakly confined atoms from the trap. First Ramsey pulse is extended to yield $P_\mathrm{dark} = 0.72$ and reduce/reverse collisional shifts. (b) Differential light shift between a high intensity test condition ($V_e=140 \textrm{ to } 550\,E_r$) and a low intensity reference condition at $V_e = 70\,E_r$. Error bars include collisional shifts and variation of trapping parameters. Markers and colors indicate lattice frequencies, lines show predictions for interpolated parameters. Results before correction for collisions ($\times$) show the reversal of the shift due to the elevated $P_\mathrm{dark}$. (c) Deviation of measured results from the model for specific trapping parameters of each measurement. (d) Sideband spectra for measurements indicated in (b) and (c) by red open diamonds. Table lists extracted trapping parameters.}
	\label{hyperpolar2}
\end{figure}

By stabilizing the lattice laser to a frequency comb referenced to an ultra-stable cavity, $\nu_\mathrm{L}$ is known to better than 100\,kHz. The parameters $\bar{n}$, $V_0$, $\zeta$ and $\delta_2$ are determined from sideband spectra typically taken both at the start and end of each experiment. To find the lattice-induced light shift, it is necessary to know the physical quantities $\nu_\mathrm{E1}$, $\tilde{\alpha}'$, $\tilde{\alpha}^{qm}$ and $\tilde{\beta}$.
Figures~\ref{hyperpolar1} and \ref{hyperpolar2} illustrate the measurements investigating $\nu_\mathrm{E1}$, $\tilde{\alpha}'$ and $\tilde{\beta}$. For this purpose, the final lattice intensity is alternated between a low intensity reference point and a high intensity test condition to determine the resulting clock frequency difference. As previously discussed, the atom number is simultaneously alternated between high and low values to extrapolate the results to zero density. The statistical uncertainty typically reaches a targeted value of $2\times 10^{-17} (\approx 10 \mathrm{\,mHz})$ for a four-hour experiment. We include the collisional shift uncertainty in this value, since its simultaneous determination avoids errors typically arising from changes in trap conditions between different experiments.

The measurements cover lattice intensities characterized by $V_0 = 90 \text{ to } 650\,E_r$, for which we find the effective depth to range from $V_e = 70\,E_r$ to $570\,E_r$. The later series of measurements includes a temporary dip in the lattice intensity, typically to $V_0^\mathrm{dip}\approx 65\,E_r$, to remove the most energetic atoms. This avoids atoms probing the outermost regions of the lattice potential, where imperfect beam overlap or the presence of higher-order radial modes may result in position-dependent variations in the intensity imbalance that cannot appropriately be handled even by the more flexible sideband-interpretation model presented here. 

\subsection{Determination of $\tilde{\alpha}^{qm}$}

In a separate measurement series, we investigate the remaining coefficient $\tilde{\alpha}^{qm}$ by determining the resulting frequency shift when alternating between vibrational states $\bar{n}\approx 1$ and $\bar{n}\approx 0$. To populate a specific vibrational state, we first apply sideband cooling to bring atoms to the $n=0$ ground state. Atoms are then excited to ${}^3P_0$ 
by a $40\, \mathrm{ms}$ clock transition pulse of $\gg \pi$ area on either the $\Delta n = 1$ blue sideband transition or the $\Delta n = 0$ carrier transition.
Clearing out the ${}^1S_0$ ground state population through a resonant pulse of the spin-polarization laser leaves an ensemble with either $\bar{n}_1 = 1.026(12)$ or $\bar{n}_0 = 0.035(8)$, averaged over all experiments. We will refer to samples prepared in this way as $[n=1]$ and $[n=0]$ in the following, and the difference in resonant frequency at $\nu_\mathrm{L} \approx \nu_\textrm{E1}$ yields $\tilde{\alpha}^{qm}$ through the term $- \tilde{\alpha}^{qm} \left(n_i+\tfrac{1}{2} \right) \sqrt{V/E_r}$  in Eq.~\ref{eq_plane_wave}.

The trap parameters for $[n=0]$ are directly determined from sideband spectra taken after the excitation pulse. Since for $[n=1]$ the evaluation of the ensemble vibrational state cannot exploit the absence of the red sideband for all atoms except a small $n>0$ population (see Appendix~\ref{sec_sideband}), we find 
it more accurate to determine the trap parameters from those of $[n=0]$ by applying corrections and uncertainties for off-resonant excitation of the carrier transition and increased trap loss of the more energetic $n=1$ atoms. For experiments performed at $V_0 = 95\,E_r$, 22\,\% of atoms are lost on excitation to $n = 1$. Due to the coupling of axial and radial vibrational energies \cite{Blatt2009} this preferentially removes atoms in higher radial motional states and thus results in an increased $\zeta_{[n=1]}$. We quantify this by repeating the experiment at lattice frequencies $\nu_\mathrm{L} = \nu_\mathrm{E1} \pm 270 \mathrm{\,MHz}$, where the term $-\tilde{\alpha}' (\nu_\mathrm{L}-\nu_\mathrm{E1})\,\zeta\,\tfrac{V_0}{E_r}$ in Eq.~\ref{eq_ensemble} dominates over the term depending on $\tilde{\alpha}^{qm}$. We then determine a difference in fractional depth $\Delta\zeta = \zeta_{[n=1]} - \zeta_{[n=0]}$  that yields the best agreement with the light shift model. For measurements at $95\,E_r$, we find $\Delta\zeta = 0.034(17)$. For measurements at $124\,E_r$, where no significant atom loss is observed on excitation, $\Delta\zeta = 0.005(9)$ is consistent with zero. The obtained values of $\Delta\zeta$ are used to determine $\zeta_{[n=1]}$  in the final evaluation. As before, measurements are performed in a multiply interleaved scheme and extrapolated to zero atom number.

\subsection{Evaluation}

We fit the combined data set for 55 experiments to the light shift model to simultaneously find the coefficients $\tilde{\alpha}'$, $\tilde{\alpha}^{qm}$, $\tilde{\beta}$ and $\nu_\mathrm{E1}$. When determining the weight of each data point, we consider the instability of the determined trapping parameters in addition to the statistical measurement uncertainties, which in turn include the extrapolation to zero density. We find a reduced $\chi^2=1.4$, and accordingly inflate the uncertainties by a factor of 1.2 \footnote{This correction is included in all uncertainties shown in figures.}.

To include systematic effects that do not average to zero over repeated measurements, we use Monte-Carlo methods to characterize their impact on the determined coefficients: The fit is repeated for numerous parameter variations, and the root-mean-square deviation from the originally determined coefficients is included in the uncertainty. Effects handled in this way include model-dependencies of $\bar{n}$ and $V_e = \zeta V_0$ extracted from the sideband analysis.

In the experiments reported here, the return beam of the lattice is attenuated to 83\,\% intensity, or a relative electric field amplitude of $a_r=0.91(4)$. It is then straightforward to calculate $r=(1+a_r )^2/ 4a_r = 1.0024(23)$. However, a change in lattice focal position discovered after the conclusion of the experiments may have affected the later series of measurements, possibly resulting in a larger than expected intensity imbalance. We investigate this by fitting the experimental data with a value $r'=r+\Delta r$ for the measurements of Fig.~\ref{hyperpolar2}, performed four months after the initial experiments. The result of $\Delta r = 0.011(6)$ is consistent with the presence of a minor running wave contribution during later experiments. We therefore assign an overall uncertainty $\sigma_r = 0.013 = \sqrt{0.011^2+0.006^2}$ based on the sensitivity of this test and incorporate it through Monte Carlo variation, rejecting unphysical values of $r<1$. This most significantly affects $\tilde{\alpha}'$ and $\tilde{\beta}$, but also results in an additional uncertainty of 520\,kHz for $\nu_\mathrm{E1}$. 

The effective value of the differential hyperpolarizability $\tilde{\beta}$ depends on the lattice polarization as $\tilde{\beta}= \tilde{\beta}^\mathrm{lin}+\xi^2(\tilde{\beta}^\mathrm{circ}-\tilde{\beta}^\mathrm{lin})$, where $\tilde{\beta}^\mathrm{lin}$ and $\tilde{\beta}^\mathrm{circ}$ are the coefficients for linear and circular lattice polarizations, and the degree of circular polarization $\xi$ relates to ellipticity angle $\chi$ as $\xi=\sin 2\chi$~\cite{Katori_Ovsiannikov2015}. As discussed in Appendix~\ref{sec_ellipticity}, we include an uncertainty of $4.6\%$ for $\tilde{\beta}$ representing an ellipticity $\chi \le 0.026\, \pi$ resulting from imperfect polarization and viewport birefringence.

Table~\ref{tab_unc} lists the extracted parameter values and their uncertainties.

\begin{table*}[!tbp]
  \caption{Results and uncertainties for the experimentally determined light shift parameters.  \label{tab_unc}}
  \begin{tabularx}{17.8 cm}{ 
      l 
      >{\raggedleft}p{0.95cm} >{\raggedleft}p{0.75cm}
      p{0.7cm} >{\raggedleft}p{0.95cm} >{\raggedleft}p{0.75cm}
      p{0.7cm} >{\raggedleft}p{1.15cm} >{\raggedleft}p{0.75cm} 
      p{0.7cm} >{\raggedleft}p{0.95cm} >{\raggedleft}p{0.75cm}
      l 
    }
    \toprule
    & \multicolumn{2}{c}{ $\tilde{\alpha}'/h$ } 
    && \multicolumn{2}{c}{ $\tilde{\alpha}^{qm}/h$ } 
    && \multicolumn{2}{c}{ $\tilde{\beta}/h$ } 
    && \multicolumn{2}{c}{ $\nu_\mathrm{E1}$ } & \\
    & \multicolumn{2}{c}{ $(\mathrm{\mu Hz / MHz})$ } 
    && \multicolumn{2}{c}{ $(\mathrm{\mu Hz})$ } 
    && \multicolumn{2}{c}{ $(\mathrm{\mu Hz})$ } 
    & \multicolumn{4}{r}{ $(+394\,798\,000\,\mathrm{MHz})$ }   \\
    contribution & value & unc. && value & unc. && value & unc. && value & unc. & \\
    \hline
    statistical uncertainty &      & 0.30 &&             & 372 &&               & 0.055 &&            & 1.26 & \\
    uncertainty of $r$            && 0.31 &&             &   65 &&               & 0.030 &&            & 0.52 &\\
    uncertainty of parameters&& 0.33 &&              &  32 &&               & 0.030 &&            & 0.02 &\\
    lattice polarization           &&         &&              &       &&               & 0.055 &&            &        & \\ 
    \hline
    overall                 & 25.74 & 0.54 && $-1027$ & 378 && $-1.194$ & 0.089 && 261.06 & 1.37 &\\
    \toprule
  \end{tabularx}
\end{table*}

\section{Discussion}
\label{sec_discussion}
It is readily apparent that the accurate determination of the fractional depth $\zeta$ is crucial to the evaluation of the lattice-induced frequency shifts. We perform an additional series of experiments to investigate the frequency shifts that result from changes to the loading procedure. Here we vary the fractional depth by loading the lattice at an intensity corresponding to either $V_0^\mathrm{cold} \approx 110\,E_r$ or $V_0^\mathrm{hot} \approx 600\,E_r$ with no successive reduction. For this hot state, $\zeta^\mathrm{hot}=0.52$ (with $\delta_2^\mathrm{hot}=0.047$) corresponds to a radial temperature $T_\rho = 30\, \mathrm{\mu K}$, while for the cold state, we find $\zeta^\mathrm{cold}=0.84$ ($\delta_2^\mathrm{cold}=0.006$) and $T_\rho = 9.2\, \mathrm{\mu K}$ after a ramp to full lattice intensity. Appendix~\ref{sec_sideband} shows the evaluation of these sideband spectra in more detail. 

\begin{figure}[!tbp]
	\includegraphics[width=8.4 cm]{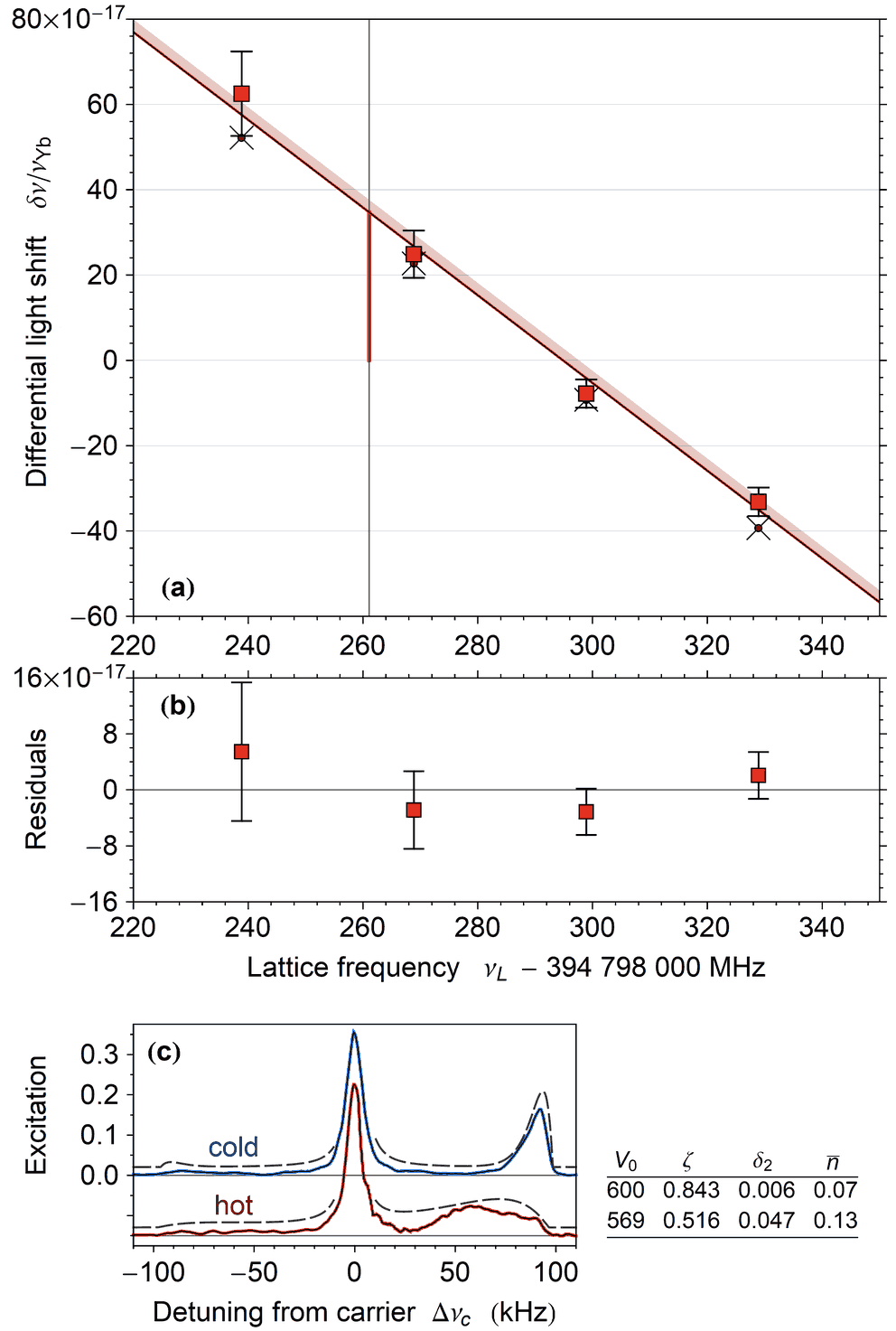}
	\caption{Test of extracted fractional depth. (a) Atoms are prepared either in a radially cold state by loading at low intensity followed by an increase in lattice depth, or in a hot state by preparation at a constant high intensity. Data points indicate the difference in resonant frequency between cold ($\zeta^\mathrm{cold}=0.843$) and hot ($\zeta^\mathrm{hot}=0.516$) state as a function of lattice frequency. Lattice intensity during interrogation is kept identical, with $V_0 \approx 600\,E_r$. Results agree with model predictions (solid line). Shaded region indicates change in model prediction for a best fit with $\zeta^\mathrm{hot}$ and $\delta_2^\mathrm{hot}$ as free parameters (see main text). Error bars include uncertainty of trapping parameters and extrapolation to zero atom number, and $\times$ indicate results before collisional correction. A fractional differential shift of $3\times 10^{-16}$ with change in radial temperature occurs even near $\nu_\mathrm{E1}$ (vertical line), largely due to hyperpolarizability. (b) Deviation from light shift model predictions. (c) Sideband spectra at full lattice depth after loading at $V_0 \approx 110\,E_r$ (cold) or $V_0\approx 600\,E_r$ (hot). Table lists extracted trapping parameters. Note that sideband cooling remains effective in reducing the axial vibrational state to near $\bar{n}=0$, but does not affect the radial temperature. Dashed lines (offset for visibility) show that a fit with the previously used model based on~\cite{Blatt2009} fails to reproduce the shape of the spectrum for energetic atoms.}
	\label{radial}
\end{figure}

Although atoms are interrogated at identical $V_0 \approx 600\,E_r$, this results in a differential frequency shift of $3\times10^{-16}$, even in the vicinity of the E1-magic frequency. As shown in Fig.~\ref{radial}, the observed frequency shifts are in excellent agreement with the predictions of the light shift model, with a reduced $\chi^2=0.46$. A fit for a hypothetical deviation of $\zeta^\mathrm{hot}$ and $\delta_2^\mathrm{hot}$ from the values obtained by sideband analysis yields best agreement for negligible corrections $\Delta\zeta = 0.012(30)$ and $\Delta\delta_2 = 0.014(27)$. 

During measurements we typically employ an adiabatic increase in lattice depth to ensure strong atomic confinement and a stable value of $\zeta$. We also find that a 9.5\,ms Doppler-cooling pulse from a pair of 556\,nm beams orthogonal to the lattice axis, controlled independently from the lasers of the magneto-optical trap, helps achieve lower radial motional states when loading the lattice at high intensity (Fig.~\ref{hyperpolar2}(a)). 

An alternative approach to the direct determination of the fractional depth $\zeta$ is to assume a fixed relation of the atomic motional state to the lattice depth, and characterize the light shifts by a number of empirical parameters that represent a specific preparation sequence \cite{Brown2017}. Within the framework presented here, these assumptions correspond to constant values of $\zeta$ and $\delta_2$, along with $\bar{n} = b_n \sqrt{V_0/E_r} - \tfrac{1}{2}$. A proportionality constant $b_n=0.03$ reproduces an axial vibrational temperature $T_z = (\bar{n}+\tfrac{1}{2}) h f_z / k_\mathrm{B} \approx 5\,\mathrm{\mu K}$ at $V_0 = 1200\,E_r$ as presented in \cite{Brown2017}. Although we find it problematic that this yields $\bar{n}<0$ for common values of $V_0\approx 100\,E_r$, it is then straightforward to recast Eq.~\ref{eq_ensemble} in the form
$\Delta\nu_\mathrm{en} / \nu_\mathrm{Yb} = -\frac{\partial\alpha^*}{\partial\nu_l}  (\nu_l - \nu_\mathrm{zero}) (V_0/E_r) - \beta^* (V_0/E_r)^2$. We find agreement with the reported values $\frac{\partial\alpha^*}{\partial\nu_l} = 2.46(10)\times 10^{-20}\,\mathrm{MHz^{-1}}$ and $\beta^* = -5.5(2)\times 10^{-22}$ for trapping parameters $\zeta = 0.516(24)$, $\delta_2 = -0.006(42)$, consistent with our observations when employing the same strategy of loading at the full lattice depth with no procedure to control the radial motional state (shown for $V_0 \approx 600\,E_r$ in Fig.~\ref{radial}). We obtain $\nu^x_\mathrm{zero} = 394,798,262.8(1.5)\,\mathrm{MHz}$ for the lattice frequency with vanishing linear lattice depth dependence, close to the reported value of $\nu_\mathrm{zero} = 394,798,267(1)\,\mathrm{MHz}$. However, for our experimentally determined value of $\tilde{\alpha}^{qm}$, we find a required correction $\nu_\mathrm{E1}-\nu_\mathrm{zero} = -1.76\,\mathrm{MHz}$ due to the contribution of the E2/M1 polarizability to the term linear in $V_0$ when considering $\bar{n} \propto \sqrt{V_0}$. This is several times larger than anticipated in \cite{Brown2017}.

To facilitate comparisons with independent measurements or new calculated values, Table~\ref{tab_coeff} lists the coefficients extracted here in conventional units as in \cite{Katori_Ovsiannikov2015}, with intensities representing a single lattice beam. To convert the units, a value of $4 \times \alpha_\mathrm{E1} = 186(7) \, \text{atomic units}$ $(h \times 34.8\,\mathrm{kHz/(kW/cm^2)})$ has been used for the electric dipole polarizability of the clock states in the vicinity of the magic frequency, according to \cite{Dzuba2010} and the uncertainty elaborated in \cite{Derevianko2011}. All values agree well with our previous results when considering the larger uncertainties of the earlier measurements. The table also includes the result of theoretical calculations \cite{
Ovsiannikov2016} and the values reported in \cite{Brown2017} as atomic properties, including corrections for thermal effects. However, these corrections seem to be underestimated when comparing the results with the values of $\tilde{\alpha}'$ and $\tilde{\beta}$ determined here. It is noteworthy that this does not affect the light shift evaluation in \cite{Brown2017}, which is based on the empirical coefficients determined directly for the motional state of the atomic samples encountered in clock interrogation. The discrepancy between the theoretically calculated value for $\tilde{\alpha}^{qm}$ and our experimental results warrants further investigation as a significant contribution to the overall clock uncertainty, while $\tilde{\alpha}'$ may act as a convenient test of calculated polarizabilities and thermal correction models.

\begin{table*}[!tbp]
  \caption{Light shift coefficients in conventional units, compared to values reported previously and results of other groups. $\mathbf{Bold}$ numbers indicate experimental results, $\mathit{italics}$ indicate theoretical calculations.\label{tab_coeff}}
  \begin{tabularx}{17.8 cm}{cp{3.8 cm}rrrr} 
    \toprule
    coeff.
    &equivalent in \cite{Katori_Ovsiannikov2015}&this work&previous~\cite{Nemitz2016}&from~\cite{Brown2017}&from~\cite{Ovsiannikov2016}\\
    \hline
    $\tilde{\alpha}'$ 
    & $\tfrac{\partial \Delta \alpha^\mathrm{E1}}{\partial \nu} / h \Big(\tfrac{\mathrm{mHz / MHz}}{\mathrm{kW / cm^2}}\Big)$
    & $\mathbf{0.443(20)}$ & $\mathbf{0.369(97)}$ & $\mathbf{0.357}$ & $\mathit{0.720}$\\
    $\tilde{\alpha}^{qm}$ 
    & $\Delta\alpha_m^{qm} / h \ \Big( \tfrac{\mathrm{mHz}}{\mathrm{kW/cm^2}} \Big)$
    & $\mathbf{-17.7(6.6)}$ & $\mathbf{-11.8(12.3)}$ & $\mathit{1.4(1.4)}$ & $\mathit{-8.06}$\\
    $\tilde{\beta}$	
    & $\Delta\beta_m^l / h \ \ \Big( \tfrac{\mathrm{\mu Hz}}{\left(\mathrm{kW/cm^2}\right)^2} \Big)$
    & $\mathbf{-354(39)}$ & $\mathbf{-552(247)}$ & $\mathbf{-153.8}$ & $\mathit{-312}$\\
    $\nu_\mathrm{E1}$ 
    & $\nu_\mathrm{E1} (\mathrm{MHz}),\ 394,798,000 +$
    & $\mathbf{261.1(1.4)}$ & $\mathbf{265.0(9.5)}$ & $\mathbf{266.6(1.0)}$ &\\
    \toprule
  \end{tabularx}
\end{table*}

\section{Conclusion}

By using the methods described here to characterize the trapping conditions, the coefficients $\tilde{\alpha}'$, $\tilde{\alpha}^{qm}$ and $\tilde{\beta}$
are directly applicable to any clock based on ${}^{171}\mathrm{Yb}$. This will allow improved accuracies, particularly for experimental designs that cannot access a sufficient range of intensities to distinguish and characterize the frequency shifts originating from hyperpolarizability. It is advisable to re-determine the magic frequency, as the apparent value is easily affected not only by beam imbalance, but also by the spectral composition of the lattice laser \cite{Pizzocaro2017, Brown2017}.

\begin{figure}[!btp]
	\includegraphics[width= 8.6 cm]{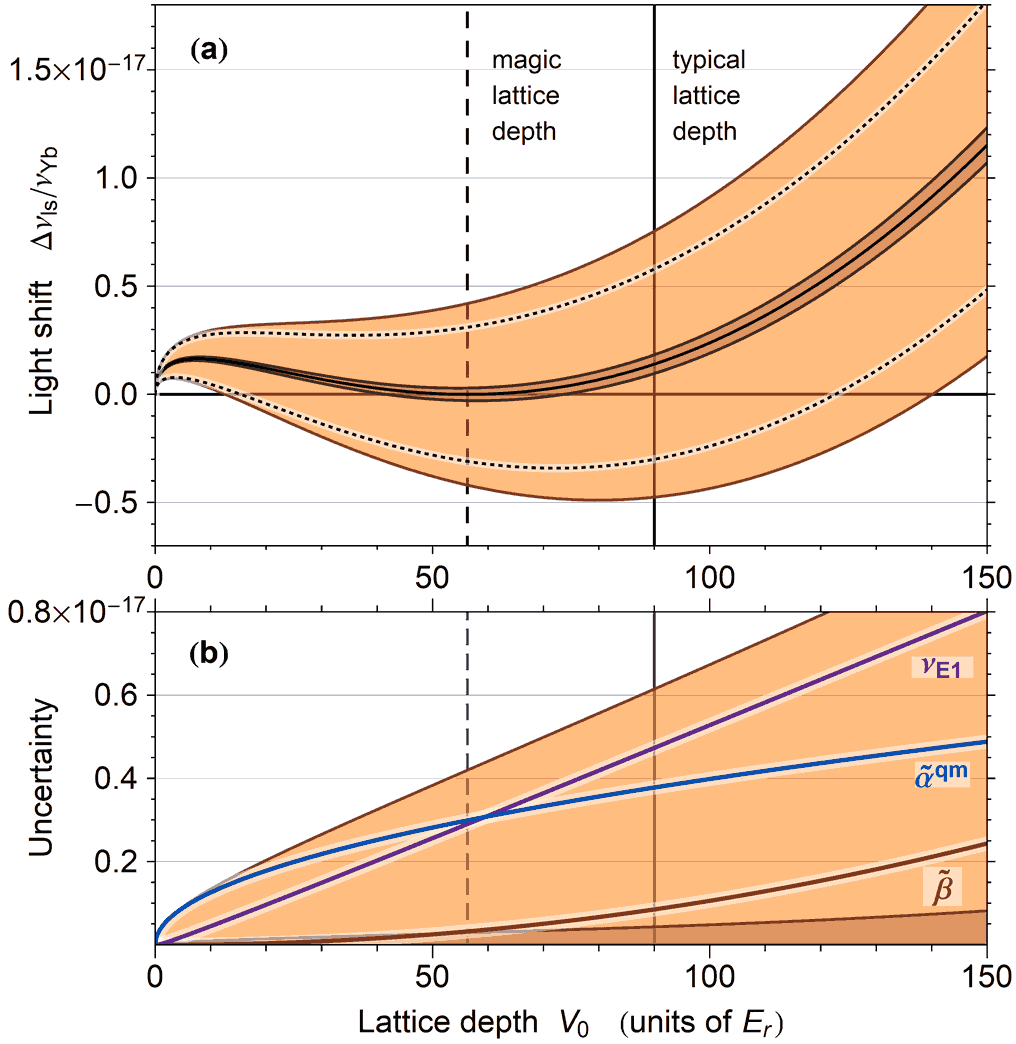}
	\caption{Modelled light shift and uncertainty. (a) Model predictions and $1\sigma$ uncertainty bands for typical trapping parameters as a function of lattice depth at the operational magic frequency $\nu_\mathrm{L} = 394,798,267\,\mathrm{MHz}$, where both the light shift and its slope become zero for an operational magic lattice depth $V_0 = 56\,E_r$ (dashed vertical line). Inner, brown-shaded region indicates uncertainties due to the determination of trap parameters. Outer, orange-shaded region indicates overall uncertainty. Dotted lines represent the overall uncertainty if covariances of coefficients are included in the evaluation. Solid vertical line indicates typical lattice depth of $V_0 = 90\,E_r$ in cryogenic operation of the Yb clock, where the model gives a fractional uncertainty of $\sigma_\mathrm{ls} = 6.1\times 10^{-18}$. (b) Breakdown of overall uncertainty, with marked contributions from $\tilde{\alpha}^{qm}$, $\tilde{\beta}$ and $\nu_\mathrm{E1}$.}
	\label{uncertainties}
\end{figure}

When applied to the typical operating conditions of our Yb optical lattice clock in its cryogenic mode of operation, characterized by $\zeta = 0.83(1)$, $\delta_2 = 0.006(2)$ and $\bar{n} = 0.10(1)$, we find a cancellation of the light shift as well as its derivative with regards to the lattice depth for magic operating conditions \cite{Katori_Ovsiannikov2015, Ushijima2018} $V_0 = 56\,E_r$ and $\nu_\mathrm{L} = 394,798,267\,\mathrm{MHz}$. Although our current lattice geometry, tilted $15^\circ$ from vertical, provides insufficient confinement against gravity
\footnote{
The peak axial restoring force $F_z=-2 \pi V_0 / \lambda$ counteracts gravity at a negligible $V_0 = 0.25 \, E_r$. The peak radial force $F_\rho = -2 V_0 /(\sqrt{e}\, w)$ requires $V_0= 120 \, E_r$ for a horizontal lattice of radius $w=70\, \mathrm{\mu m}$. For a $15^\circ$ tilt, $F_\rho = m_\mathrm{Yb} \, g \, \sin{15^\circ}$ requires a minimum of $V_0 = 31 \, E_r$, and usable numbers of trapped atoms are observed only for $V_0 \ge 60 \, E_r$.
}, targeting such an operational magic lattice depth will allow future optical lattice clocks to further minimize lattice-induced light shifts and the resulting uncertainties. The evaluation presented here supports an uncertainty of $4.2\times 10^{-18}$, as shown in Fig.~\ref{uncertainties}, which gives an overview of the light shifts as a function of lattice depth and a breakdown of the uncertainty contributions.

At the value of $V_0 = 90\,E_r$ recently used in cryogenic operation, the residual fractional uncertainty is $\sigma_\mathrm{ls} = 6.1\times 10^{-18}$ if we make the simplifying assumption of errors that are uncorrelated between coefficients. A treatment incorporating the full covariance matrix obtained from the fit of the measured data shows that the correlations reduce the overall uncertainty to $4.4\times 10^{-18}$. The contributions from the uncertainties of the trapping parameters and from the uncertainty of the hyperpolarizability coefficient $\tilde{\beta}$ are only $5\times 10^{-19}$ and $9\times 10^{-19}$ respectively. Significant further improvement is therefore possible through additional measurements that require less lattice intensity and can be performed in the cryogenic configuration of the clock.

The results presented here will enable our cryogenic Yb optical lattice clock to operate with a systematic uncertainty of only a few times $10^{-18}$, clearing the path towards measurements of clock frequency ratios with an uncertainty of $\le 10^{-17}$.

\appendix
\section{Uncertainty of trapping parameters}
To evaluate the uncertainty originating from the trapping parameters, we consider two contributions: The first is a statistical uncertainty that represents both changes in the actual trapping conditions, and the repeatability of the parameter determination from sideband spectra in the presence of measurement noise. We determine this value by comparing multiple sets of trap parameters extracted from spectra taken for the same conditions, typically at the beginning and end of a measurement. The second contribution is an estimate of the systematic uncertainty. For the effective depth $V_e$, this is based on multiple evaluations of the same set of sideband spectra under varied assumptions for interrogation time and probe intensity. For the ensemble vibrational state $\bar{n}$, we consider the effects of populations in vibrational states $n>1$ that are not included in the sideband model. The systematic uncertainty for the quadratic correction $\delta_2$ is based on the measurements varying the atomic temperature, shown in fig.~\ref{radial}. Table~\ref{tab_parameters} lists the results for both reproducibility and systematic uncertainty together with nominal values for typical clock operating conditions in cryogenic configuration. 

\begin{table}[!tbh]
  \caption{Parameter uncertainties.\label{tab_parameters}}
  \begin{ruledtabular}
  \begin{tabular}{lcccc}
    parameter & & nominal & reproducibility & syst. unc. \\
    \hline
    lattice depth  & $V_0$ & $ 90\,E_r$ & $0.035 \times V_0$ & \\	
    fractional depth & $\zeta$ & $0.83$ & $0.012$ & \\	
    effective depth & $V_e$ & $75\,E_r$ & $0.030\times V_e$ & $0.012 \times V_e$ \\
    quadratic correction & $\delta_2$ & $0.006$ & $0.060 \times \delta_2$ & $0.3 \times \delta_2$ \\
    vibrational state & $\bar{n}$ & $0.10$ & $0.031$ & $0.013$ \\
  \end{tabular}
  \end{ruledtabular}
\end{table}

\section{Sideband evaluation}
\label{sec_sideband}
\begin{figure}[!tbp]
	\includegraphics[width= 8.5 cm]{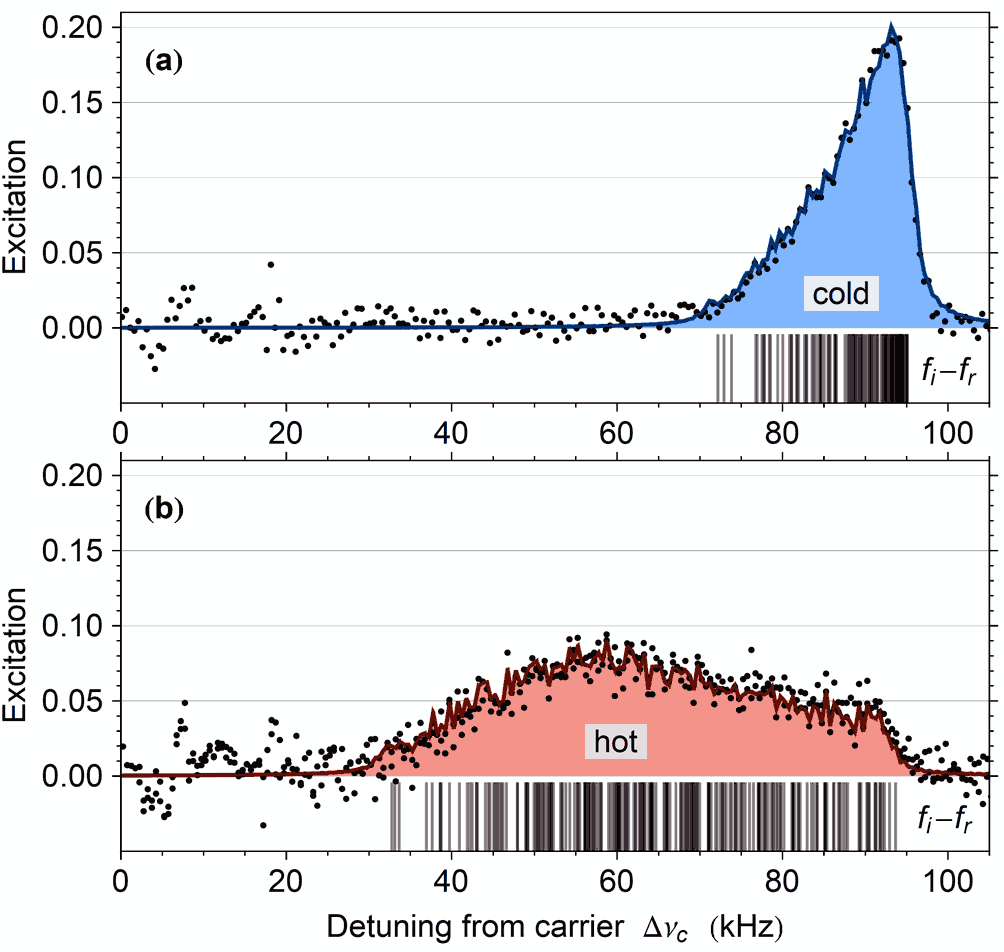}
	\caption{Reconstruction of sideband spectra for $T=1 \, \mathrm{ms}$ pulses. Points show excitation for blue $n\to n+1$ sideband after subtraction of the carrier transition at $\Delta\nu_c = 0\,\mathrm{kHz}$. Solid lines and shaded area give the numerically reconstructed spectrum for an optimized set of trap frequencies $f_i$, shown below as vertical lines at $f_i-f_r^\mathrm{lat}$, indicating their contribution to the spectrum. (a) Cold sample prepared by loading at reduced lattice intensity, followed by adiabatic ramp to $V_0 = 597\,E_r$. Spectrum reconstructed using $111\,f_i$ values. (b) Hot sample prepared by directly loading the lattice at $V_0 = 597\,E_r$. Spectrum reconstructed using $222\,f_i$ values.}
	\label{sideband_fit}
\end{figure}

In the first step of the sideband evaluation,
a Lorentzian fit of half-width at half-maximum $\gamma$ yields the Rabi frequency $f_c$ ($=\Omega_c/ 2\pi$) of the (dephased) $n \to n$ carrier transition as
$f_c = \gamma \approx 2.5\,\mathrm{kHz}$. The fit is then subtracted from the data, and we primarily investigate the $n \to n+1$ (blue) sideband, which is present for atoms in all axial vibrational states. Figure~\ref{sideband_fit} shows exemplary data for the 'hot' and 'cold' states of section~\ref{sec_discussion}.
After a pulse of duration $T$ applied to a singular atom in a sinusoidal axial potential characterized by trap frequency $f_i$, the sideband transition spectrum at a detuning $\Delta\nu_c$ from the carrier is described by an excitation
\begin{equation}
\label{eq_sideband_spectrum}
P^\mathrm{sb}_i = \frac{f_\mathrm{sb}^2}{f_\mathrm{sb}^2+(\Delta\nu_c - \delta^B_i )^2} \sin^2 \left( \pi \sqrt{f_\mathrm{sb}^2+(\Delta\nu_c - \delta^B_i)^2}\, T \right)\ ,
\end{equation}
where $f_\mathrm{sb}$ is the sideband Rabi frequency and $\delta^B_i = f_i - (n_i+1)  f_r^\mathrm{lat}$ is the sideband transition frequency (see Eq.~\ref{eq_delta_BSB}). We will refer to $f_r^\mathrm{lat} = E_r / h = 2024\,\mathrm{Hz}$ as the lattice photon recoil frequency. The equivalent clock photon recoil frequency is $f_r^\mathrm{clk} = 3489\,\mathrm{Hz}$, and using this, the Lamb-Dicke parameter can be written as $\eta = \sqrt{f_r^\mathrm{clk} / f_i}$, ranging from 0.18 to 0.35 over the relevant range of trap frequencies. The $n-1 \leftrightarrow n$ sideband transitions are then excited with Rabi frequency $f_\mathrm{sb} \approx \sqrt{n} \, \eta \, f_c$ (see \cite{Wineland1979} for a complete expression). We ignore dephasing effects since at low pulse areas only the central feature near $\delta^B_i $ contributes significantly to the spectrum, calculated as the sum 
\begin{equation}
P^\mathrm{sb}(\Delta\nu_c) = a \frac{1}{N} \sum_i{ P_i^\mathrm{sb}(\Delta\nu_c)}\ .
\end{equation}
To address errors in assumed line shape and $f_c$, factor $a \approx 0.5 \textrm{ to } 2$ is adjusted to match the spectral data.

A number $N$ of $f_i$, sufficient to obtain a smooth $P^\mathrm{sb}$, is selected in the frequency range where significant excitation probability is observed. The $f_i$ are iteratively adjusted until the sum of squared residuals from the observed spectrum has converged on a minimum. While a fixed set of $f_i$ with adjustable weights would allow a more efficient algorithm, this tends to fit local noise features outside the spectrum.

The sideband transition frequencies $\delta^B_i$ are unambiguous for a known vibrational state $n$ throughout the ensemble, which is ensured by sideband cooling to $n=0$. To include a residual $n \ge 0$ population, we calculate a hypothetical spectrum for $n = 1 \to n = 0$ using the same $f_i$. The relative magnitude of the observed red sideband (which is absent for $n=0$) yields an $n=1$ contribution of typically less than 10\% (see Fig.~\ref{radial}c). We include the admixture of $n=1$ in the calculation of the blue sideband and refine the $f_i$ accordingly. Fig.~\ref{sideband_fit} shows reconstructed spectra.

\begin{figure}[!btp]
	\includegraphics[width= 8.5 cm]{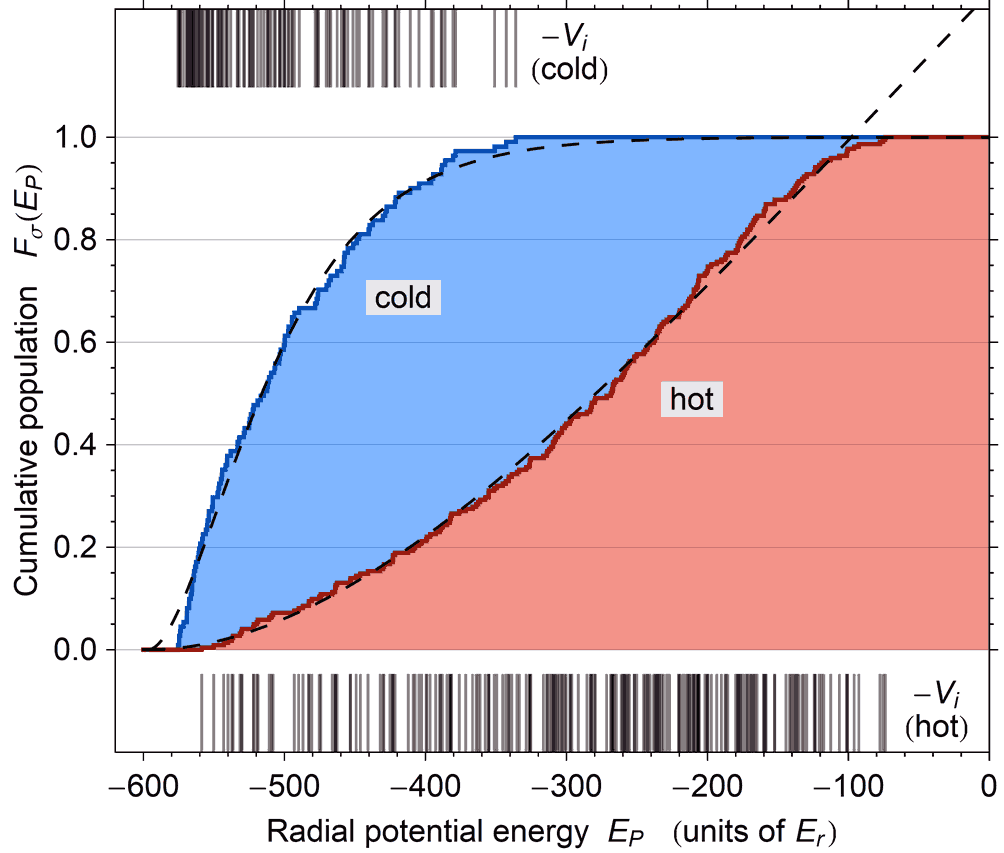}
	\caption{Cumulative population distributions for hot ($T_\rho = 30 \, \mathrm{\mu K}$) and cold ($T_\rho = 9.2 \, \mathrm{\mu K}$) samples. Solid, colored lines indicate fractional population integrated over discretized $\sigma_d(E_{P})$, starting from $\hat{E} = -V_0$. Vertical lines above and below indicate contributions $E_{P,i} = -V_i$. Dashed lines represent the thermal model given in the text.}
	\label{cumulative_dist}
\end{figure}

The extracted set of $f_i$ provides a discretized approximation of $\sigma(V)$ as a series of delta-functions:
\begin{equation}
\sigma_d(V) = \frac{1}{N}\sum_i{\delta(V-V_i)}\ \mathrm{,\ \ with \ } V_i = \frac{h^2  f_i^2}{4 \, E_r}
\end{equation}
To compare the results to a thermal model, we consider $V_i$ as the axial potential depth experienced by an atom with radial potential energy $E_{P,i} = -V_i$. For a thermal ensemble of atoms at temperature $T_B$ in weak harmonic confinement with central potential $\hat{E} = -V_0$, we find the Boltzmann distribution
\begin{equation} 
\sigma_B(E_{P}) = \left(\frac{2}{k_B \, T_B}\right)^2 (E_{P}-\hat{E} )  e^{-2\frac{E_{P}-\hat{E}}{k_B \,  T_B} }  \ .
\end{equation}
The factor $( E_{P} - \hat{E} )$ represents the density-of-states in the radially symmetric potential and gives rise to the characteristic slope at the outer edge of the sideband spectrum. The factor of 2 in the exponent accounts for equal contributions of kinetic energy for the two radial degrees of freedom. To compare $\sigma_d$ and $\sigma_B$, Fig.~\ref{cumulative_dist} shows the respective cumulative distribution functions
\begin{equation}
F_\sigma(E_{P})=\int_{\hat{E}}^{E_{P}}\sigma(E'_{P}) dE'_{P} \ .
\end{equation}
The thermal energy of the cold sample was limited by loading at reduced lattice intensity, and $\sigma_d$ 
agrees well with a thermal distribution for $T_B = 9.4\,\mathrm{\mu K}$. We assign a descriptive radial temperature $T_\rho$ (available directly from the set of $V_i$) according to a potential energy contribution of $\tfrac{1}{2} k_B T$ from each of the two radial degrees of freedom:
\begin{equation}
T_\rho = \frac{\bar{E}_{P} - \hat{E}}{k_B} = \frac{V_0-V_e}{ k_B } = (1-\zeta) \frac{V_0}{k_B} \ ,
\end{equation}
where $\bar{E}_{P}=-V_e$ is the mean radial potential energy. Note that since $V_0$ represents a single $V_i$, it is more easily affected by noise in the sideband spectrum or the details of the numerical reconstruction than $V_e$, which represents the entire ensemble (see Table~\ref{tab_parameters} and $V_0$ values in Fig.~\ref{radial}).

For the cold sample, $T_\rho = 9.2\,\mathrm{\mu K}$ is in good agreement with $T_B$. 
For the hot sample resulting from loading the lattice at full intensity, $T_\rho = 30\,\mathrm{\mu K}$, and we find $\sigma_d$ 
effectively truncated to $E_{P} \le -140\,E_r$. This truncation limit invalidates the assumption of a harmonic radial potential: If the equipartition theorem were to hold true, the total radial energy of the most energetic observed atoms $2(E_{P}-\hat{E} ) \approx 900\, E_r$ significantly exceeds the trap depth of $V_0=597\, E_r$ found for combined evaluation of hot and cold spectra. Instead, the shallow outer region of the Gaussian lattice potential not only affects the encountered density of states, but also leads to a lower kinetic energy contribution at large potential energy.

The numerical model presented here is insensitive to the potential distribution since it directly relates the observed trap frequencies to potential depths and thus the resulting lattice light shifts. The same is not true for simple thermal models assuming a Boltzmann distribution in a harmonic potential, approximations that are clearly invalid for energetic atoms. However, the addition of an adjustable truncation parameter might provide a worthwhile extension of such models when the signal-to-noise ratio of the acquired data (or the mixture of axial vibrational states) make a direct numerical evaluation unfeasible.

\section{Lattice ellipticity}
\label{sec_ellipticity}

Lattice ellipticity is predominantly induced by birefringence of the upper vacuum viewport. It is characterized by an angle of ellipticity such that $\tan{\chi_b}$ gives the ratio of minor to major axis of the polarization ellipse. We set an experimental limit for this contribution based on the clock transition spectrum: The magnetic field is routinely aligned to the polarization axis of the clock laser by minimizing the $\Delta m_F=\pm 1$ contributions. The minimum obtainable Rabi frequency is $f_\sigma = f_c \tan{\chi_b}$ after accounting for Clebsch-Gordan coefficients and the decomposition into circular components. For a $3\, \pi$-pulse exciting the $\Delta m_F=0$ transition, the observed suppression to $P_\sigma =\sin^2{(2 \pi \times f_\sigma \, t / 2)} \le 0.12$ yields $\chi_b<0.024\, \pi$. We take this as an upper limit for the birefringence-induced ellipticity at the longer wavelength of the lattice laser.

Furthermore, the lattice polarizer may admit an orthogonal polarization component with intensity $I_\mathrm{ort}$ in addition to the desired component of intensity $I_\mathrm{par}$. The resulting polarization ellipse is characterized by $\tan{\chi_p} \le \sqrt{I_\mathrm{ort} / I_\mathrm{par}}$, where the largest ellipticity occurs for an orthogonal component with a relative phase $\pm \pi/2$. For $I_\mathrm{ort} / I_\mathrm{par} \le 0.001$ after the polarizer, we find $\chi_p \le 0.01\, \pi$.

Considering both independent contributions, we expect an ellipticity no larger than $\chi = 0.026\, \pi$. With a sensitivity factor $s=|\tilde{\beta}^\mathrm{circ}/\tilde{\beta}^\mathrm{lin} -1| = 1.77$, we take the value obtained for $\tilde{\beta}$ to represent $\tilde{\beta}^\mathrm{lin}$ to within a fractional error of $\delta \tilde{\beta} / \tilde{\beta} = s \, (\sin{2 \, \chi})^2 = 0.046$, which we include in the uncertainty.

\begin{acknowledgments}
This work is supported by JST ERATO Grant Number JPMJER1002 (Japan), by JSPS Grant-in-Aid for Specially Promoted Research Grant Number JP16H06284, and by the Photon Frontier Network Program of the Ministry of Education, Culture, Sports, Science and Technology, Japan.
\end{acknowledgments}

\bibliography{lightshift_arxiv_bib}

\begin{thebibliography}{33}%
\makeatletter
\providecommand \@ifxundefined [1]{%
 \@ifx{#1\undefined}
}%
\providecommand \@ifnum [1]{%
 \ifnum #1\expandafter \@firstoftwo
 \else \expandafter \@secondoftwo
 \fi
}%
\providecommand \@ifx [1]{%
 \ifx #1\expandafter \@firstoftwo
 \else \expandafter \@secondoftwo
 \fi
}%
\providecommand \natexlab [1]{#1}%
\providecommand \enquote  [1]{``#1''}%
\providecommand \bibnamefont  [1]{#1}%
\providecommand \bibfnamefont [1]{#1}%
\providecommand \citenamefont [1]{#1}%
\providecommand \href@noop [0]{\@secondoftwo}%
\providecommand \href [0]{\begingroup \@sanitize@url \@href}%
\providecommand \@href[1]{\@@startlink{#1}\@@href}%
\providecommand \@@href[1]{\endgroup#1\@@endlink}%
\providecommand \@sanitize@url [0]{\catcode `\\12\catcode `\$12\catcode
  `\&12\catcode `\#12\catcode `\^12\catcode `\_12\catcode `\%12\relax}%
\providecommand \@@startlink[1]{}%
\providecommand \@@endlink[0]{}%
\providecommand \url  [0]{\begingroup\@sanitize@url \@url }%
\providecommand \@url [1]{\endgroup\@href {#1}{\urlprefix }}%
\providecommand \urlprefix  [0]{URL }%
\providecommand \Eprint [0]{\href }%
\providecommand \doibase [0]{http://dx.doi.org/}%
\providecommand \selectlanguage [0]{\@gobble}%
\providecommand \bibinfo  [0]{\@secondoftwo}%
\providecommand \bibfield  [0]{\@secondoftwo}%
\providecommand \translation [1]{[#1]}%
\providecommand \BibitemOpen [0]{}%
\providecommand \bibitemStop [0]{}%
\providecommand \bibitemNoStop [0]{.\EOS\space}%
\providecommand \EOS [0]{\spacefactor3000\relax}%
\providecommand \BibitemShut  [1]{\csname bibitem#1\endcsname}%
\let\auto@bib@innerbib\@empty
\bibitem [{\citenamefont {Ushijima}\ \emph {et~al.}(2015)\citenamefont
  {Ushijima}, \citenamefont {Takamoto}, \citenamefont {Das}, \citenamefont
  {Ohkubo},\ and\ \citenamefont {Katori}}]{Ushijima2015}%
  \BibitemOpen
  \bibfield  {author} {\bibinfo {author} {\bibfnamefont {I.}~\bibnamefont
  {Ushijima}}, \bibinfo {author} {\bibfnamefont {M.}~\bibnamefont {Takamoto}},
  \bibinfo {author} {\bibfnamefont {M.}~\bibnamefont {Das}}, \bibinfo {author}
  {\bibfnamefont {T.}~\bibnamefont {Ohkubo}}, \ and\ \bibinfo {author}
  {\bibfnamefont {H.}~\bibnamefont {Katori}},\ }\href {\doibase
  10.1038/nphoton.2015.5} {\bibfield  {journal} {\bibinfo  {journal} {Nature
  Photon.}\ }\textbf {\bibinfo {volume} {9}},\ \bibinfo {pages} {185} (\bibinfo
  {year} {2015})}\BibitemShut {NoStop}%
\bibitem [{\citenamefont {Nicholson}\ \emph {et~al.}(2015)\citenamefont
  {Nicholson}, \citenamefont {Campbell}, \citenamefont {Hutson}, \citenamefont
  {Marti}, \citenamefont {Bloom}, \citenamefont {McNally}, \citenamefont
  {Zhang}, \citenamefont {Barrett}, \citenamefont {Safronova}, \citenamefont
  {Strouse}, \citenamefont {Tew},\ and\ \citenamefont {Ye}}]{Nicholson2015}%
  \BibitemOpen
  \bibfield  {author} {\bibinfo {author} {\bibfnamefont {T.~L.}\ \bibnamefont
  {Nicholson}}, \bibinfo {author} {\bibfnamefont {S.~L.}\ \bibnamefont
  {Campbell}}, \bibinfo {author} {\bibfnamefont {R.~B.}\ \bibnamefont
  {Hutson}}, \bibinfo {author} {\bibfnamefont {G.~E.}\ \bibnamefont {Marti}},
  \bibinfo {author} {\bibfnamefont {B.~J.}\ \bibnamefont {Bloom}}, \bibinfo
  {author} {\bibfnamefont {R.~L.}\ \bibnamefont {McNally}}, \bibinfo {author}
  {\bibfnamefont {W.}~\bibnamefont {Zhang}}, \bibinfo {author} {\bibfnamefont
  {M.~D.}\ \bibnamefont {Barrett}}, \bibinfo {author} {\bibfnamefont {M.~S.}\
  \bibnamefont {Safronova}}, \bibinfo {author} {\bibfnamefont {G.~F.}\
  \bibnamefont {Strouse}}, \bibinfo {author} {\bibfnamefont {W.~L.}\
  \bibnamefont {Tew}}, \ and\ \bibinfo {author} {\bibfnamefont
  {J.}~\bibnamefont {Ye}},\ }\href@noop {} {\bibfield  {journal} {\bibinfo
  {journal} {Nature Commun.}\ }\textbf {\bibinfo {volume} {6}},\ \bibinfo
  {pages} {6896} (\bibinfo {year} {2015})}\BibitemShut {NoStop}%
\bibitem [{\citenamefont {Huntemann}\ \emph {et~al.}(2016)\citenamefont
  {Huntemann}, \citenamefont {Sanner}, \citenamefont {Lipphardt}, \citenamefont
  {Tamm},\ and\ \citenamefont {Peik}}]{Huntemann2016}%
  \BibitemOpen
  \bibfield  {author} {\bibinfo {author} {\bibfnamefont {N.}~\bibnamefont
  {Huntemann}}, \bibinfo {author} {\bibfnamefont {C.}~\bibnamefont {Sanner}},
  \bibinfo {author} {\bibfnamefont {B.}~\bibnamefont {Lipphardt}}, \bibinfo
  {author} {\bibfnamefont {C.}~\bibnamefont {Tamm}}, \ and\ \bibinfo {author}
  {\bibfnamefont {E.}~\bibnamefont {Peik}},\ }\href {\doibase
  10.1103/PhysRevLett.116.063001} {\bibfield  {journal} {\bibinfo  {journal}
  {Phys. Rev. Lett.}\ }\textbf {\bibinfo {volume} {116}},\ \bibinfo {pages}
  {063001} (\bibinfo {year} {2016})}\BibitemShut {NoStop}%
\bibitem [{\citenamefont {McGrew}\ \emph {et~al.}(2018)\citenamefont {McGrew},
  \citenamefont {Zhang}, \citenamefont {Fasano}, \citenamefont {Schäffer},
  \citenamefont {Beloy}, \citenamefont {Nicolodi}, \citenamefont {Brown},
  \citenamefont {Hinkley}, \citenamefont {Milani}, \citenamefont {Schioppo},
  \citenamefont {Yoon},\ and\ \citenamefont {Ludlow}}]{McGrew2018}%
  \BibitemOpen
  \bibfield  {author} {\bibinfo {author} {\bibfnamefont {W.~F.}\ \bibnamefont
  {McGrew}}, \bibinfo {author} {\bibfnamefont {X.}~\bibnamefont {Zhang}},
  \bibinfo {author} {\bibfnamefont {R.~J.}\ \bibnamefont {Fasano}}, \bibinfo
  {author} {\bibfnamefont {S.~A.}\ \bibnamefont {Schäffer}}, \bibinfo {author}
  {\bibfnamefont {K.}~\bibnamefont {Beloy}}, \bibinfo {author} {\bibfnamefont
  {D.}~\bibnamefont {Nicolodi}}, \bibinfo {author} {\bibfnamefont {R.~C.}\
  \bibnamefont {Brown}}, \bibinfo {author} {\bibfnamefont {N.}~\bibnamefont
  {Hinkley}}, \bibinfo {author} {\bibfnamefont {G.}~\bibnamefont {Milani}},
  \bibinfo {author} {\bibfnamefont {M.}~\bibnamefont {Schioppo}}, \bibinfo
  {author} {\bibfnamefont {T.~H.}\ \bibnamefont {Yoon}}, \ and\ \bibinfo
  {author} {\bibfnamefont {A.~D.}\ \bibnamefont {Ludlow}},\ }\href {\doibase
  10.1038/s41586-018-0738-2} {\bibfield  {journal} {\bibinfo  {journal}
  {Nature}\ }\textbf {\bibinfo {volume} {564}},\ \bibinfo {pages} {87}
  (\bibinfo {year} {2018})}\BibitemShut {NoStop}%
\bibitem [{\citenamefont {Katori}\ \emph {et~al.}(2003)\citenamefont {Katori},
  \citenamefont {Takamoto}, \citenamefont {Pal'chikov},\ and\ \citenamefont
  {Ovsiannikov}}]{Katori2003}%
  \BibitemOpen
  \bibfield  {author} {\bibinfo {author} {\bibfnamefont {H.}~\bibnamefont
  {Katori}}, \bibinfo {author} {\bibfnamefont {M.}~\bibnamefont {Takamoto}},
  \bibinfo {author} {\bibfnamefont {V.~G.}\ \bibnamefont {Pal'chikov}}, \ and\
  \bibinfo {author} {\bibfnamefont {V.~D.}\ \bibnamefont {Ovsiannikov}},\
  }\href {\doibase 10.1103/PhysRevLett.91.173005} {\bibfield  {journal}
  {\bibinfo  {journal} {Phys. Rev. Lett.}\ }\textbf {\bibinfo {volume} {91}},\
  \bibinfo {pages} {173005} (\bibinfo {year} {2003})}\BibitemShut {NoStop}%
\bibitem [{\citenamefont {Taichenachev}\ \emph {et~al.}(2008)\citenamefont
  {Taichenachev}, \citenamefont {Yudin}, \citenamefont {Ovsiannikov},
  \citenamefont {Pal'chikov},\ and\ \citenamefont {Oates}}]{Taichenachev2008}%
  \BibitemOpen
  \bibfield  {author} {\bibinfo {author} {\bibfnamefont {A.~V.}\ \bibnamefont
  {Taichenachev}}, \bibinfo {author} {\bibfnamefont {V.~I.}\ \bibnamefont
  {Yudin}}, \bibinfo {author} {\bibfnamefont {V.~D.}\ \bibnamefont
  {Ovsiannikov}}, \bibinfo {author} {\bibfnamefont {V.~G.}\ \bibnamefont
  {Pal'chikov}}, \ and\ \bibinfo {author} {\bibfnamefont {C.~W.}\ \bibnamefont
  {Oates}},\ }\href {\doibase 10.1103/PhysRevLett.101.193601} {\bibfield
  {journal} {\bibinfo  {journal} {Phys. Rev. Lett.}\ }\textbf {\bibinfo
  {volume} {101}},\ \bibinfo {pages} {193601} (\bibinfo {year}
  {2008})}\BibitemShut {NoStop}%
\bibitem [{\citenamefont {Ovsiannikov}\ \emph {et~al.}(2013)\citenamefont
  {Ovsiannikov}, \citenamefont {Pal'chikov}, \citenamefont {Taichenachev},
  \citenamefont {Yudin},\ and\ \citenamefont {Katori}}]{Ovsiannikov2013}%
  \BibitemOpen
  \bibfield  {author} {\bibinfo {author} {\bibfnamefont {V.~D.}\ \bibnamefont
  {Ovsiannikov}}, \bibinfo {author} {\bibfnamefont {V.~G.}\ \bibnamefont
  {Pal'chikov}}, \bibinfo {author} {\bibfnamefont {A.~V.}\ \bibnamefont
  {Taichenachev}}, \bibinfo {author} {\bibfnamefont {V.~I.}\ \bibnamefont
  {Yudin}}, \ and\ \bibinfo {author} {\bibfnamefont {H.}~\bibnamefont
  {Katori}},\ }\href {\doibase 10.1103/PhysRevA.88.013405} {\bibfield
  {journal} {\bibinfo  {journal} {Phys. Rev. A}\ }\textbf {\bibinfo {volume}
  {88}},\ \bibinfo {pages} {013405} (\bibinfo {year} {2013})}\BibitemShut
  {NoStop}%
\bibitem [{\citenamefont {Ovsiannikov}\ \emph {et~al.}(2016)\citenamefont
  {Ovsiannikov}, \citenamefont {Mormo}, \citenamefont {Palchikov},\ and\
  \citenamefont {Katori}}]{Ovsiannikov2016}%
  \BibitemOpen
  \bibfield  {author} {\bibinfo {author} {\bibfnamefont {V.~D.}\ \bibnamefont
  {Ovsiannikov}}, \bibinfo {author} {\bibfnamefont {S.~I.}\ \bibnamefont
  {Mormo}}, \bibinfo {author} {\bibfnamefont {V.~G.}\ \bibnamefont
  {Palchikov}}, \ and\ \bibinfo {author} {\bibfnamefont {H.}~\bibnamefont
  {Katori}},\ }\href {\doibase 10.1103/PhysRevA.93.043420} {\bibfield
  {journal} {\bibinfo  {journal} {Phys. Rev. A}\ }\textbf {\bibinfo {volume}
  {93}},\ \bibinfo {pages} {043420} (\bibinfo {year} {2016})}\BibitemShut
  {NoStop}%
\bibitem [{\citenamefont {Brusch}\ \emph {et~al.}(2006)\citenamefont {Brusch},
  \citenamefont {Le~Targat}, \citenamefont {Baillard}, \citenamefont
  {Fouch\'e},\ and\ \citenamefont {Lemonde}}]{Brusch2006}%
  \BibitemOpen
  \bibfield  {author} {\bibinfo {author} {\bibfnamefont {A.}~\bibnamefont
  {Brusch}}, \bibinfo {author} {\bibfnamefont {R.}~\bibnamefont {Le~Targat}},
  \bibinfo {author} {\bibfnamefont {X.}~\bibnamefont {Baillard}}, \bibinfo
  {author} {\bibfnamefont {M.}~\bibnamefont {Fouch\'e}}, \ and\ \bibinfo
  {author} {\bibfnamefont {P.}~\bibnamefont {Lemonde}},\ }\href {\doibase
  10.1103/PhysRevLett.96.103003} {\bibfield  {journal} {\bibinfo  {journal}
  {Phys. Rev. Lett.}\ }\textbf {\bibinfo {volume} {96}},\ \bibinfo {pages}
  {103003} (\bibinfo {year} {2006})}\BibitemShut {NoStop}%
\bibitem [{\citenamefont {Barber}\ \emph {et~al.}(2008)\citenamefont {Barber},
  \citenamefont {Stalnaker}, \citenamefont {Lemke}, \citenamefont {Poli},
  \citenamefont {Oates}, \citenamefont {Fortier}, \citenamefont {Diddams},
  \citenamefont {Hollberg},\ and\ \citenamefont {Hoyt}}]{Barber2008}%
  \BibitemOpen
  \bibfield  {author} {\bibinfo {author} {\bibfnamefont {Z.~W.}\ \bibnamefont
  {Barber}}, \bibinfo {author} {\bibfnamefont {J.~E.}\ \bibnamefont
  {Stalnaker}}, \bibinfo {author} {\bibfnamefont {N.~D.}\ \bibnamefont
  {Lemke}}, \bibinfo {author} {\bibfnamefont {N.}~\bibnamefont {Poli}},
  \bibinfo {author} {\bibfnamefont {C.~W.}\ \bibnamefont {Oates}}, \bibinfo
  {author} {\bibfnamefont {T.~M.}\ \bibnamefont {Fortier}}, \bibinfo {author}
  {\bibfnamefont {S.~A.}\ \bibnamefont {Diddams}}, \bibinfo {author}
  {\bibfnamefont {L.}~\bibnamefont {Hollberg}}, \ and\ \bibinfo {author}
  {\bibfnamefont {C.~W.}\ \bibnamefont {Hoyt}},\ }\href {\doibase
  10.1103/PhysRevLett.100.103002} {\bibfield  {journal} {\bibinfo  {journal}
  {Phys. Rev. Lett}\ }\textbf {\bibinfo {volume} {100}},\ \bibinfo {pages}
  {103002} (\bibinfo {year} {2008})}\BibitemShut {NoStop}%
\bibitem [{\citenamefont {Westergaard}\ \emph {et~al.}(2011)\citenamefont
  {Westergaard}, \citenamefont {Lodewyck}, \citenamefont {Lorini},
  \citenamefont {Lecallier}, \citenamefont {Burt}, \citenamefont {Zawada},
  \citenamefont {Millo},\ and\ \citenamefont {Lemonde}}]{Westergaard2011}%
  \BibitemOpen
  \bibfield  {author} {\bibinfo {author} {\bibfnamefont {P.~G.}\ \bibnamefont
  {Westergaard}}, \bibinfo {author} {\bibfnamefont {J.}~\bibnamefont
  {Lodewyck}}, \bibinfo {author} {\bibfnamefont {L.}~\bibnamefont {Lorini}},
  \bibinfo {author} {\bibfnamefont {A.}~\bibnamefont {Lecallier}}, \bibinfo
  {author} {\bibfnamefont {E.~A.}\ \bibnamefont {Burt}}, \bibinfo {author}
  {\bibfnamefont {M.}~\bibnamefont {Zawada}}, \bibinfo {author} {\bibfnamefont
  {J.}~\bibnamefont {Millo}}, \ and\ \bibinfo {author} {\bibfnamefont
  {P.}~\bibnamefont {Lemonde}},\ }\href {\doibase
  10.1103/PhysRevLett.106.210801} {\bibfield  {journal} {\bibinfo  {journal}
  {Phys. Rev. Lett.}\ }\textbf {\bibinfo {volume} {106}},\ \bibinfo {pages}
  {210801} (\bibinfo {year} {2011})}\BibitemShut {NoStop}%
\bibitem [{\citenamefont {Brown}\ \emph {et~al.}(2017)\citenamefont {Brown},
  \citenamefont {Phillips}, \citenamefont {Beloy}, \citenamefont {McGrew},
  \citenamefont {Schioppo}, \citenamefont {Fasano}, \citenamefont {Milani},
  \citenamefont {Zhang}, \citenamefont {Hinkley}, \citenamefont {Leopardi},
  \citenamefont {Yoon}, \citenamefont {Nicolodi}, \citenamefont {Fortier},\
  and\ \citenamefont {Ludlow}}]{Brown2017}%
  \BibitemOpen
  \bibfield  {author} {\bibinfo {author} {\bibfnamefont {R.~C.}\ \bibnamefont
  {Brown}}, \bibinfo {author} {\bibfnamefont {N.~B.}\ \bibnamefont {Phillips}},
  \bibinfo {author} {\bibfnamefont {K.}~\bibnamefont {Beloy}}, \bibinfo
  {author} {\bibfnamefont {W.~F.}\ \bibnamefont {McGrew}}, \bibinfo {author}
  {\bibfnamefont {M.}~\bibnamefont {Schioppo}}, \bibinfo {author}
  {\bibfnamefont {R.~J.}\ \bibnamefont {Fasano}}, \bibinfo {author}
  {\bibfnamefont {G.}~\bibnamefont {Milani}}, \bibinfo {author} {\bibfnamefont
  {X.}~\bibnamefont {Zhang}}, \bibinfo {author} {\bibfnamefont
  {N.}~\bibnamefont {Hinkley}}, \bibinfo {author} {\bibfnamefont
  {H.}~\bibnamefont {Leopardi}}, \bibinfo {author} {\bibfnamefont {T.~H.}\
  \bibnamefont {Yoon}}, \bibinfo {author} {\bibfnamefont {D.}~\bibnamefont
  {Nicolodi}}, \bibinfo {author} {\bibfnamefont {T.~M.}\ \bibnamefont
  {Fortier}}, \ and\ \bibinfo {author} {\bibfnamefont {A.~D.}\ \bibnamefont
  {Ludlow}},\ }\href {\doibase 10.1103/PhysRevLett.119.253001} {\bibfield
  {journal} {\bibinfo  {journal} {Phys. Rev. Lett.}\ }\textbf {\bibinfo
  {volume} {119}},\ \bibinfo {pages} {253001} (\bibinfo {year}
  {2017})}\BibitemShut {NoStop}%
\bibitem [{\citenamefont {Ushijima}\ \emph {et~al.}(2018)\citenamefont
  {Ushijima}, \citenamefont {Takamoto},\ and\ \citenamefont
  {Katori}}]{Ushijima2018}%
  \BibitemOpen
  \bibfield  {author} {\bibinfo {author} {\bibfnamefont {I.}~\bibnamefont
  {Ushijima}}, \bibinfo {author} {\bibfnamefont {M.}~\bibnamefont {Takamoto}},
  \ and\ \bibinfo {author} {\bibfnamefont {H.}~\bibnamefont {Katori}},\ }\href
  {\doibase 10.1103/PhysRevLett.121.263202} {\bibfield  {journal} {\bibinfo
  {journal} {Phys. Rev. Lett.}\ }\textbf {\bibinfo {volume} {121}},\ \bibinfo
  {pages} {263202} (\bibinfo {year} {2018})}\BibitemShut {NoStop}%
\bibitem [{\citenamefont {Nemitz}\ \emph {et~al.}(2016)\citenamefont {Nemitz},
  \citenamefont {Ohkubo}, \citenamefont {Takamoto}, \citenamefont {Ushijima},
  \citenamefont {Das}, \citenamefont {Ohmae},\ and\ \citenamefont
  {Katori}}]{Nemitz2016}%
  \BibitemOpen
  \bibfield  {author} {\bibinfo {author} {\bibfnamefont {N.}~\bibnamefont
  {Nemitz}}, \bibinfo {author} {\bibfnamefont {T.}~\bibnamefont {Ohkubo}},
  \bibinfo {author} {\bibfnamefont {M.}~\bibnamefont {Takamoto}}, \bibinfo
  {author} {\bibfnamefont {I.}~\bibnamefont {Ushijima}}, \bibinfo {author}
  {\bibfnamefont {M.}~\bibnamefont {Das}}, \bibinfo {author} {\bibfnamefont
  {N.}~\bibnamefont {Ohmae}}, \ and\ \bibinfo {author} {\bibfnamefont
  {H.}~\bibnamefont {Katori}},\ }\href {\doibase 10.1038/nphoton.2016.20}
  {\bibfield  {journal} {\bibinfo  {journal} {Nature Photon.}\ }\textbf
  {\bibinfo {volume} {10}},\ \bibinfo {pages} {258} (\bibinfo {year}
  {2016})}\BibitemShut {NoStop}%
\bibitem [{\citenamefont {Le~Targat}\ \emph {et~al.}(2013)\citenamefont
  {Le~Targat}, \citenamefont {Lorini}, \citenamefont {Le~Coq}, \citenamefont
  {Zawada}, \citenamefont {Guéna}, \citenamefont {Abgrall}, \citenamefont
  {Gurov}, \citenamefont {Rosenbusch}, \citenamefont {Rovera}, \citenamefont
  {Nagórny}, \citenamefont {Gartman}, \citenamefont {Westergaard},
  \citenamefont {Tobar}, \citenamefont {Lours}, \citenamefont {Santarelli},
  \citenamefont {Clairon}, \citenamefont {Bize}, \citenamefont {Laurent},
  \citenamefont {Lemonde},\ and\ \citenamefont {Lodewyck}}]{LeTargat2013}%
  \BibitemOpen
  \bibfield  {author} {\bibinfo {author} {\bibfnamefont {R.}~\bibnamefont
  {Le~Targat}}, \bibinfo {author} {\bibfnamefont {L.}~\bibnamefont {Lorini}},
  \bibinfo {author} {\bibfnamefont {Y.}~\bibnamefont {Le~Coq}}, \bibinfo
  {author} {\bibfnamefont {M.}~\bibnamefont {Zawada}}, \bibinfo {author}
  {\bibfnamefont {J.}~\bibnamefont {Guéna}}, \bibinfo {author} {\bibfnamefont
  {M.}~\bibnamefont {Abgrall}}, \bibinfo {author} {\bibfnamefont
  {M.}~\bibnamefont {Gurov}}, \bibinfo {author} {\bibfnamefont
  {P.}~\bibnamefont {Rosenbusch}}, \bibinfo {author} {\bibfnamefont {D.~G.}\
  \bibnamefont {Rovera}}, \bibinfo {author} {\bibfnamefont {B.}~\bibnamefont
  {Nagórny}}, \bibinfo {author} {\bibfnamefont {R.}~\bibnamefont {Gartman}},
  \bibinfo {author} {\bibfnamefont {P.~G.}\ \bibnamefont {Westergaard}},
  \bibinfo {author} {\bibfnamefont {M.~E.}\ \bibnamefont {Tobar}}, \bibinfo
  {author} {\bibfnamefont {M.}~\bibnamefont {Lours}}, \bibinfo {author}
  {\bibfnamefont {G.}~\bibnamefont {Santarelli}}, \bibinfo {author}
  {\bibfnamefont {A.}~\bibnamefont {Clairon}}, \bibinfo {author} {\bibfnamefont
  {S.}~\bibnamefont {Bize}}, \bibinfo {author} {\bibfnamefont {P.}~\bibnamefont
  {Laurent}}, \bibinfo {author} {\bibfnamefont {P.}~\bibnamefont {Lemonde}}, \
  and\ \bibinfo {author} {\bibfnamefont {J.}~\bibnamefont {Lodewyck}},\ }\href
  {\doibase 10.1038/ncomms3109} {\bibfield  {journal} {\bibinfo  {journal}
  {Nature Commun.}\ }\textbf {\bibinfo {volume} {4}},\ \bibinfo {pages} {2109}
  (\bibinfo {year} {2013})}\BibitemShut {NoStop}%
\bibitem [{\citenamefont {Yamanaka}\ \emph {et~al.}(2015)\citenamefont
  {Yamanaka}, \citenamefont {Ohmae}, \citenamefont {Ushijima}, \citenamefont
  {Takamoto},\ and\ \citenamefont {Katori}}]{Yamanaka2015}%
  \BibitemOpen
  \bibfield  {author} {\bibinfo {author} {\bibfnamefont {K.}~\bibnamefont
  {Yamanaka}}, \bibinfo {author} {\bibfnamefont {N.}~\bibnamefont {Ohmae}},
  \bibinfo {author} {\bibfnamefont {I.}~\bibnamefont {Ushijima}}, \bibinfo
  {author} {\bibfnamefont {M.}~\bibnamefont {Takamoto}}, \ and\ \bibinfo
  {author} {\bibfnamefont {H.}~\bibnamefont {Katori}},\ }\href {\doibase
  10.1103/PhysRevLett.114.230801} {\bibfield  {journal} {\bibinfo  {journal}
  {Phys. Rev. Lett}\ }\textbf {\bibinfo {volume} {114}},\ \bibinfo {pages}
  {230801} (\bibinfo {year} {2015})}\BibitemShut {NoStop}%
\bibitem [{\citenamefont {Mukaiyama}\ \emph {et~al.}(2003)\citenamefont
  {Mukaiyama}, \citenamefont {Katori}, \citenamefont {Ido}, \citenamefont
  {Li},\ and\ \citenamefont {Kuwata-Gonokami}}]{Mukaiyama2003}%
  \BibitemOpen
  \bibfield  {author} {\bibinfo {author} {\bibfnamefont {T.}~\bibnamefont
  {Mukaiyama}}, \bibinfo {author} {\bibfnamefont {H.}~\bibnamefont {Katori}},
  \bibinfo {author} {\bibfnamefont {T.}~\bibnamefont {Ido}}, \bibinfo {author}
  {\bibfnamefont {Y.}~\bibnamefont {Li}}, \ and\ \bibinfo {author}
  {\bibfnamefont {M.}~\bibnamefont {Kuwata-Gonokami}},\ }\href {\doibase
  10.1103/PhysRevLett.90.113002} {\bibfield  {journal} {\bibinfo  {journal}
  {Phys. Rev. Lett.}\ }\textbf {\bibinfo {volume} {90}},\ \bibinfo {pages}
  {113002} (\bibinfo {year} {2003})}\BibitemShut {NoStop}%
\bibitem [{\citenamefont {Zhang}\ \emph {et~al.}(2015)\citenamefont {Zhang},
  \citenamefont {Zhou}, \citenamefont {Chen}, \citenamefont {Gao},
  \citenamefont {Han}, \citenamefont {Yao}, \citenamefont {Xu}, \citenamefont
  {Li}, \citenamefont {Xu}, \citenamefont {Jiang}, \citenamefont {Bi},
  \citenamefont {Ma},\ and\ \citenamefont {Xu}}]{Zhang2015}%
  \BibitemOpen
  \bibfield  {author} {\bibinfo {author} {\bibfnamefont {X.}~\bibnamefont
  {Zhang}}, \bibinfo {author} {\bibfnamefont {M.}~\bibnamefont {Zhou}},
  \bibinfo {author} {\bibfnamefont {N.}~\bibnamefont {Chen}}, \bibinfo {author}
  {\bibfnamefont {Q.}~\bibnamefont {Gao}}, \bibinfo {author} {\bibfnamefont
  {C.}~\bibnamefont {Han}}, \bibinfo {author} {\bibfnamefont {Y.}~\bibnamefont
  {Yao}}, \bibinfo {author} {\bibfnamefont {P.}~\bibnamefont {Xu}}, \bibinfo
  {author} {\bibfnamefont {S.}~\bibnamefont {Li}}, \bibinfo {author}
  {\bibfnamefont {Y.}~\bibnamefont {Xu}}, \bibinfo {author} {\bibfnamefont
  {Y.}~\bibnamefont {Jiang}}, \bibinfo {author} {\bibfnamefont
  {Z.}~\bibnamefont {Bi}}, \bibinfo {author} {\bibfnamefont {L.}~\bibnamefont
  {Ma}}, \ and\ \bibinfo {author} {\bibfnamefont {X.}~\bibnamefont {Xu}},\
  }\href {\doibase 10.1088/1612-2011/12/2/025501} {\bibfield  {journal}
  {\bibinfo  {journal} {Laser Phys. Lett.}\ }\textbf {\bibinfo {volume} {12}},\
  \bibinfo {pages} {025501} (\bibinfo {year} {2015})}\BibitemShut {NoStop}%
\bibitem [{\citenamefont {Pizzocaro}\ \emph {et~al.}(2017)\citenamefont
  {Pizzocaro}, \citenamefont {Thoumany}, \citenamefont {Rauf}, \citenamefont
  {Bregolin}, \citenamefont {Milani}, \citenamefont {Clivati}, \citenamefont
  {Costanzo}, \citenamefont {Levi},\ and\ \citenamefont
  {Calonico}}]{Pizzocaro2017}%
  \BibitemOpen
  \bibfield  {author} {\bibinfo {author} {\bibfnamefont {M.}~\bibnamefont
  {Pizzocaro}}, \bibinfo {author} {\bibfnamefont {P.}~\bibnamefont {Thoumany}},
  \bibinfo {author} {\bibfnamefont {B.}~\bibnamefont {Rauf}}, \bibinfo {author}
  {\bibfnamefont {F.}~\bibnamefont {Bregolin}}, \bibinfo {author}
  {\bibfnamefont {G.}~\bibnamefont {Milani}}, \bibinfo {author} {\bibfnamefont
  {C.}~\bibnamefont {Clivati}}, \bibinfo {author} {\bibfnamefont {G.~A.}\
  \bibnamefont {Costanzo}}, \bibinfo {author} {\bibfnamefont {F.}~\bibnamefont
  {Levi}}, \ and\ \bibinfo {author} {\bibfnamefont {D.}~\bibnamefont
  {Calonico}},\ }\href {\doibase 10.1088/1681-7575/aa4e62} {\bibfield
  {journal} {\bibinfo  {journal} {Metrologia}\ }\textbf {\bibinfo {volume}
  {54}},\ \bibinfo {pages} {102} (\bibinfo {year} {2017})}\BibitemShut
  {NoStop}%
\bibitem [{\citenamefont {Kim}\ \emph {et~al.}(2017)\citenamefont {Kim},
  \citenamefont {Heo}, \citenamefont {Lee}, \citenamefont {Park}, \citenamefont
  {Hong}, \citenamefont {Hwang},\ and\ \citenamefont {Yu}}]{Kim2017}%
  \BibitemOpen
  \bibfield  {author} {\bibinfo {author} {\bibfnamefont {H.}~\bibnamefont
  {Kim}}, \bibinfo {author} {\bibfnamefont {M.-S.}\ \bibnamefont {Heo}},
  \bibinfo {author} {\bibfnamefont {W.-K.}\ \bibnamefont {Lee}}, \bibinfo
  {author} {\bibfnamefont {C.~Y.}\ \bibnamefont {Park}}, \bibinfo {author}
  {\bibfnamefont {H.-G.}\ \bibnamefont {Hong}}, \bibinfo {author}
  {\bibfnamefont {S.-W.}\ \bibnamefont {Hwang}}, \ and\ \bibinfo {author}
  {\bibfnamefont {D.-H.}\ \bibnamefont {Yu}},\ }\href {\doibase
  10.7567/JJAP.56.050302} {\bibfield  {journal} {\bibinfo  {journal} {Jap. J.
  Appl. Phys.}\ }\textbf {\bibinfo {volume} {56}},\ \bibinfo {pages} {050302}
  (\bibinfo {year} {2017})}\BibitemShut {NoStop}%
\bibitem [{\citenamefont {Dzuba}\ and\ \citenamefont
  {Derevianko}(2010)}]{Dzuba2010}%
  \BibitemOpen
  \bibfield  {author} {\bibinfo {author} {\bibfnamefont {V.~A.}\ \bibnamefont
  {Dzuba}}\ and\ \bibinfo {author} {\bibfnamefont {A.}~\bibnamefont
  {Derevianko}},\ }\href {\doibase 10.1088/0953-4075/43/7/074011} {\bibfield
  {journal} {\bibinfo  {journal} {J. Phys. B: At. Mol. Opt. Phys.}\ }\textbf
  {\bibinfo {volume} {43}},\ \bibinfo {pages} {074011} (\bibinfo {year}
  {2010})}\BibitemShut {NoStop}%
\bibitem [{\citenamefont {Ludlow}\ \emph {et~al.}(2011)\citenamefont {Ludlow},
  \citenamefont {Lemke}, \citenamefont {Sherman},\ and\ \citenamefont
  {Oates}}]{Ludlow2011}%
  \BibitemOpen
  \bibfield  {author} {\bibinfo {author} {\bibfnamefont {A.~D.}\ \bibnamefont
  {Ludlow}}, \bibinfo {author} {\bibfnamefont {N.~D.}\ \bibnamefont {Lemke}},
  \bibinfo {author} {\bibfnamefont {J.~A.}\ \bibnamefont {Sherman}}, \ and\
  \bibinfo {author} {\bibfnamefont {C.~W.}\ \bibnamefont {Oates}},\ }\href
  {\doibase 10.1103/PhysRevA.84.052724} {\bibfield  {journal} {\bibinfo
  {journal} {Phys. Rev. A}\ }\textbf {\bibinfo {volume} {84}},\ \bibinfo
  {pages} {052724} (\bibinfo {year} {2011})}\BibitemShut {NoStop}%
\bibitem [{\citenamefont {Yanagimoto}\ \emph {et~al.}(2018)\citenamefont
  {Yanagimoto}, \citenamefont {Nemitz}, \citenamefont {Bregolin},\ and\
  \citenamefont {Katori}}]{Yanagimoto2018}%
  \BibitemOpen
  \bibfield  {author} {\bibinfo {author} {\bibfnamefont {R.}~\bibnamefont
  {Yanagimoto}}, \bibinfo {author} {\bibfnamefont {N.}~\bibnamefont {Nemitz}},
  \bibinfo {author} {\bibfnamefont {F.}~\bibnamefont {Bregolin}}, \ and\
  \bibinfo {author} {\bibfnamefont {H.}~\bibnamefont {Katori}},\ }\href
  {\doibase 10.1103/PhysRevA.98.012704} {\bibfield  {journal} {\bibinfo
  {journal} {Phys. Rev. A}\ }\textbf {\bibinfo {volume} {98}},\ \bibinfo
  {pages} {012704} (\bibinfo {year} {2018})}\BibitemShut {NoStop}%
\bibitem [{\citenamefont {Kanjilal}\ and\ \citenamefont
  {Blume}(2004)}]{Kanjilal2004}%
  \BibitemOpen
  \bibfield  {author} {\bibinfo {author} {\bibfnamefont {K.}~\bibnamefont
  {Kanjilal}}\ and\ \bibinfo {author} {\bibfnamefont {D.}~\bibnamefont
  {Blume}},\ }\href {\doibase 10.1103/PhysRevA.70.042709} {\bibfield  {journal}
  {\bibinfo  {journal} {Phys. Rev. A}\ }\textbf {\bibinfo {volume} {70}},\
  \bibinfo {pages} {042709} (\bibinfo {year} {2004})}\BibitemShut {NoStop}%
\bibitem [{\citenamefont {Lemke}\ \emph {et~al.}(2011)\citenamefont {Lemke},
  \citenamefont {von Stecher}, \citenamefont {Sherman}, \citenamefont {Rey},
  \citenamefont {Oates},\ and\ \citenamefont {Ludlow}}]{Lemke2011}%
  \BibitemOpen
  \bibfield  {author} {\bibinfo {author} {\bibfnamefont {N.~D.}\ \bibnamefont
  {Lemke}}, \bibinfo {author} {\bibfnamefont {J.}~\bibnamefont {von Stecher}},
  \bibinfo {author} {\bibfnamefont {J.~A.}\ \bibnamefont {Sherman}}, \bibinfo
  {author} {\bibfnamefont {A.~M.}\ \bibnamefont {Rey}}, \bibinfo {author}
  {\bibfnamefont {C.~W.}\ \bibnamefont {Oates}}, \ and\ \bibinfo {author}
  {\bibfnamefont {A.~D.}\ \bibnamefont {Ludlow}},\ }\href {\doibase
  10.1103/PhysRevLett.107.103902} {\bibfield  {journal} {\bibinfo  {journal}
  {Phys. Rev. Lett.}\ }\textbf {\bibinfo {volume} {107}},\ \bibinfo {pages}
  {103902} (\bibinfo {year} {2011})}\BibitemShut {NoStop}%
\bibitem [{\citenamefont {Dörscher}\ \emph {et~al.}(2018)\citenamefont
  {Dörscher}, \citenamefont {Schwarz}, \citenamefont {Al-Masoudi},
  \citenamefont {Falke}, \citenamefont {Sterr},\ and\ \citenamefont
  {Lisdat}}]{Doerscher2018}%
  \BibitemOpen
  \bibfield  {author} {\bibinfo {author} {\bibfnamefont {S.}~\bibnamefont
  {Dörscher}}, \bibinfo {author} {\bibfnamefont {R.}~\bibnamefont {Schwarz}},
  \bibinfo {author} {\bibfnamefont {A.}~\bibnamefont {Al-Masoudi}}, \bibinfo
  {author} {\bibfnamefont {S.}~\bibnamefont {Falke}}, \bibinfo {author}
  {\bibfnamefont {U.}~\bibnamefont {Sterr}}, \ and\ \bibinfo {author}
  {\bibfnamefont {C.}~\bibnamefont {Lisdat}},\ }\href {\doibase
  10.1103/PhysRevA.97.063419} {\bibfield  {journal} {\bibinfo  {journal} {Phys.
  Rev. A}\ }\textbf {\bibinfo {volume} {97}},\ \bibinfo {pages} {063419}
  (\bibinfo {year} {2018})}\BibitemShut {NoStop}%
\bibitem [{\citenamefont {Katori}\ \emph {et~al.}(2015)\citenamefont {Katori},
  \citenamefont {Ovsiannikov}, \citenamefont {Marmo},\ and\ \citenamefont
  {Palchikov}}]{Katori_Ovsiannikov2015}%
  \BibitemOpen
  \bibfield  {author} {\bibinfo {author} {\bibfnamefont {H.}~\bibnamefont
  {Katori}}, \bibinfo {author} {\bibfnamefont {V.~D.}\ \bibnamefont
  {Ovsiannikov}}, \bibinfo {author} {\bibfnamefont {S.~I.}\ \bibnamefont
  {Marmo}}, \ and\ \bibinfo {author} {\bibfnamefont {V.~G.}\ \bibnamefont
  {Palchikov}},\ }\href {\doibase 10.1103/PhysRevA.91.052503} {\bibfield
  {journal} {\bibinfo  {journal} {Phys. Rev. A}\ }\textbf {\bibinfo {volume}
  {91}},\ \bibinfo {pages} {052503} (\bibinfo {year} {2015})}\BibitemShut
  {NoStop}%
\bibitem [{\citenamefont {Blatt}\ \emph {et~al.}(2009)\citenamefont {Blatt},
  \citenamefont {Thomsen}, \citenamefont {Campbell}, \citenamefont {Ludlow},
  \citenamefont {Swallows}, \citenamefont {Martin}, \citenamefont {Boyd},\ and\
  \citenamefont {Ye}}]{Blatt2009}%
  \BibitemOpen
  \bibfield  {author} {\bibinfo {author} {\bibfnamefont {S.}~\bibnamefont
  {Blatt}}, \bibinfo {author} {\bibfnamefont {J.~W.}\ \bibnamefont {Thomsen}},
  \bibinfo {author} {\bibfnamefont {G.~K.}\ \bibnamefont {Campbell}}, \bibinfo
  {author} {\bibfnamefont {A.~D.}\ \bibnamefont {Ludlow}}, \bibinfo {author}
  {\bibfnamefont {M.~D.}\ \bibnamefont {Swallows}}, \bibinfo {author}
  {\bibfnamefont {M.~J.}\ \bibnamefont {Martin}}, \bibinfo {author}
  {\bibfnamefont {M.~M.}\ \bibnamefont {Boyd}}, \ and\ \bibinfo {author}
  {\bibfnamefont {J.}~\bibnamefont {Ye}},\ }\href {\doibase
  10.1103/PhysRevA.80.052703} {\bibfield  {journal} {\bibinfo  {journal} {Phys.
  Rev. A}\ }\textbf {\bibinfo {volume} {80}},\ \bibinfo {pages} {052703}
  (\bibinfo {year} {2009})}\BibitemShut {NoStop}%
\bibitem [{Note1()}]{Note1}%
  \BibitemOpen
  \bibinfo {note} {Note that we use a slightly different notation here, such
  that $(\zeta + \delta _m)^m = \zeta _m^\protect \mathrm {Sr}$}\BibitemShut
  {NoStop}%
\bibitem [{Note2()}]{Note2}%
  \BibitemOpen
  \bibinfo {note} {This correction is included in all uncertainties shown in
  figures.}\BibitemShut {Stop}%
\bibitem [{\citenamefont {Derevianko}\ and\ \citenamefont
  {Porsev}(2011)}]{Derevianko2011}%
  \BibitemOpen
  \bibfield  {author} {\bibinfo {author} {\bibfnamefont {A.}~\bibnamefont
  {Derevianko}}\ and\ \bibinfo {author} {\bibfnamefont {S.~G.}\ \bibnamefont
  {Porsev}},\ }in\ \href {\doibase
  https://doi.org/10.1016/B978-0-12-385508-4.00009-7} {\emph {\bibinfo
  {booktitle} {Advances in Atomic, Molecular, and Optical Physics}}},\ \bibinfo
  {series} {Advances In Atomic, Molecular, and Optical Physics}, Vol.~\bibinfo
  {volume} {60},\ \bibinfo {editor} {edited by\ \bibinfo {editor}
  {\bibfnamefont {E.}~\bibnamefont {Arimondo}}, \bibinfo {editor}
  {\bibfnamefont {P.}~\bibnamefont {Berman}}, \ and\ \bibinfo {editor}
  {\bibfnamefont {C.}~\bibnamefont {Lin}}}\ (\bibinfo  {publisher} {Academic
  Press},\ \bibinfo {year} {2011})\ pp.\ \bibinfo {pages} {415 --
  459}\BibitemShut {NoStop}%
\bibitem [{Note3()}]{Note3}%
  \BibitemOpen
  \bibinfo {note} {The peak axial restoring force $F_z=-2 \pi V_0 / \lambda $
  counteracts gravity at a negligible $V_0 = 0.25 \protect \tmspace
  +\thinmuskip {.1667em} E_r$. The peak radial force $F_\rho = -2 V_0
  /(\protect \sqrt {e}\protect \tmspace +\thinmuskip {.1667em} w)$ requires
  $V_0= 120 \protect \tmspace +\thinmuskip {.1667em} E_r$ for a horizontal
  lattice of radius $w=70\protect \tmspace +\thinmuskip {.1667em} \protect
  \mathrm {\mu m}$. For a $15^\circ $ tilt, $F_\rho = m_\protect \mathrm {Yb}
  \protect \tmspace +\thinmuskip {.1667em} g \protect \tmspace +\thinmuskip
  {.1667em} \protect \qopname \relax o{sin}{15^\circ }$ requires a minimum of
  $V_0 = 31 \protect \tmspace +\thinmuskip {.1667em} E_r$, and usable numbers
  of trapped atoms are observed only for $V_0 \ge 60 \protect \tmspace
  +\thinmuskip {.1667em} E_r$.}\BibitemShut {Stop}%
\bibitem [{\citenamefont {Wineland}\ and\ \citenamefont
  {Itano}(1979)}]{Wineland1979}%
  \BibitemOpen
  \bibfield  {author} {\bibinfo {author} {\bibfnamefont {D.~J.}\ \bibnamefont
  {Wineland}}\ and\ \bibinfo {author} {\bibfnamefont {W.~M.}\ \bibnamefont
  {Itano}},\ }\href {\doibase 10.1103/PhysRevA.20.1521} {\bibfield  {journal}
  {\bibinfo  {journal} {Phys. Rev. A}\ }\textbf {\bibinfo {volume} {20}},\
  \bibinfo {pages} {1521} (\bibinfo {year} {1979})}\BibitemShut {NoStop}%
\end{thebibliography}%

\end{document}